\documentclass[preprint,1p,times,authoryear]{elsarticle}
\usepackage{hyperref}

\usepackage{graphicx} 
\graphicspath{{Pictures/}}

\usepackage{caption}
\usepackage{multirow}
\usepackage{subcaption} 
\usepackage{booktabs} 
\usepackage{amsmath} 
\usepackage{amssymb}
\usepackage{color}
\usepackage{enumitem}
\journal{International Journal of Plasticity}

\biboptions{authoryear}

\begin{document}

\begin{frontmatter}
\title{Applied Machine Learning to Predict Stress Hotspots I: Face Centered Cubic Materials}%
\author[cmu]{Ankita Mangal}
\ead{mangalanks@gmail.com}
\author[cmu]{Elizabeth A. Holm\corref{cor1}}
\ead{eaholm@andrew.cmu.edu}

\cortext[cor1]{Corresponding author}
\address[cmu]{Department of Materials Science and Engineering, Carnegie Mellon University, 5000 Forbes Ave, Pittsburgh, PA 15213, USA}

\begin{abstract}
We investigate the formation of stress hotspots in polycrystalline materials under uniaxial tensile deformation by integrating full field crystal plasticity based deformation models and machine learning techniques to gain data driven insights about microstructural properties. Synthetic 3D microstructures are created representing single phase equiaxed microstructures for generic copper alloys. Uniaxial tensile deformation is simulated using a 3-D full-field, image-based Fast Fourier Transform (FFT) technique with rate-sensitive crystal plasticity, to get local micro-mechanical fields (stress and strain rates). Stress hotspots are defined as the grains having stress values above the 90th percentile of the stress distribution. Hotspot neighborhoods are then characterized using metrics that reflect local crystallography, geometry, and connectivity. This data is used to create input feature vectors to train a random forest learning algorithm, which predicts the grains that will become stress hotspots. We are able to achieve an area under the receiving operating characteristic curve (ROC-AUC) of 0.74 for face centered cubic materials modeled on generic copper alloys. The results show the power and the limitations of the machine learning approach applied to the polycrystalline grain networks.
\end{abstract}

\end{frontmatter}


\section{Introduction}
Ductile fracture is one of the most common modes of failure in materials and occurs by the nucleation, growth and coalescence of microscopic voids. \cite{Rimmer1959} established that these voids grow under stress by accumulating vacancies, and that void nucleation is induced by stress . \cite{Qidwai2009} show that an applied stress on a material is heterogeneously distributed between the grains, creating regions of stress accumulations, so-called stress hotspots.  Stress distribution between grains is dependent on the local microstructural features which in turn influence the location of void nucleation. We propose using machine learning techniques to study the impact of microstructural features on stress hotspots. Predicting damage nucleation is important because fracture ultimately defines the useful lifetime of a material.

Modern texture analysis techniques, such as near field and far field High Energy X-Ray Diffraction Microscopy (nf-HEDM, ff-HEDM) (\cite{Lienert2011}), have made three dimensional characterization of microstructures possible. This kind of mesoscale microstructural data, consisting of grain crystallography, centroids and strain fields, is well suited to machine learning techniques. Following this trend, \cite{Orme2016} have recently used decision trees to find insights about the driving forces behind deformation twinning in a magnesium alloy. However, stress hotspots are rare events; consisting of less than 10\% of the material volume and hence require a large dataset for statistical learning, which is not amenable to the currently existing HEDM datasets. Instead, we meet this requirement by using a simulation generated data set. Uniaxial tensile deformation is simulated in a number of synthetic microstructures using an image based full field crystal plasticity Elasto-Viscoplastic Fast Fourier Transform (EVPFFT) model (\cite{Lebensohn2012}).  Simulating deformation in materials gives us the advantage of preserving both the initial and final structures, thus enabling us to turn back the clock and correlate hot spots to initial microstructural features. The techniques developed in this work are directly transferable to HEDM datasets.

With the advent of material informatics, machine learning has been used to search for stable compounds across composition space, extract correlations between physical characteristics and observed properties and search for materials with useful properties (\cite{Orme2016,Rajan2015}). Taking inspiration from these successes, we use data mining and machine learning techniques at the mesoscale to build models which can be used to predict the probable failure locations based on known data in similar materials. These models help us gain insights about the characteristics in the local structure such as texture and geometry, that allow stress hotspots to form, and how such regions can be identified in a microstructure.

In this paper, plastic deformation of single phase face centered cubic (FCC) polycrystals is studied to ascertain the local microstructural characteristics related to the regions of high stress concentrations. A companion paper addresses similar issues in hexagonal close packed materials (\cite{Mangal2017c}). A random forest learning algorithm is chosen for its ease of use and interpretability of the machine learning framework. We use statistical descriptors of microstructures describing the crystallography (orientation distribution function, Schmid factor, misorientations) and  geometry (grain shape, grain boundary types) which have not been used earlier to understand their correlation with hotspot locations. The objective is twofold: predict probable failure locations, and identify the microstructural features that caus them, to facilitate microstructure engineering for materials design. 

\section{Methods}
\subsection{Dataset Generation}
\subsubsection{Synthetic Microstructures}

First, a dataset of synthetic microstructure images is built using Dream.3D (\cite{Groeber2014}). Synthetic equiaxed polycrystalline microstructures with about 5000 grains each and a mean grain size of 2.7 microns are created for a set of six representative textures in FCC materials shown in Figure \ref{fig:FCCtex}. For each texture kind (eg. Goss, Basal fiber etc.), six microstructure instantiations are created, thus resulting in about 30000 grains per crystallographic texture. The texture intensity is characterized by multiples of random density (MRD) which is the intensity of a crystallite orientation with respect to it's intensity in a randomly textured material. For each texture studied here, the texture intensity is varied from weakly textured ($<$10 MRD) to strongly textured ($>$30 MRD) among the six instantiations. The microstructures are discretized on a 128x128x128 grid, which allows the use of image based crystal plasticity models. 

The stress distribution in a microstructure is highly dependent on the crystal system and the texture. Hence we keep the crystal system constant (FCC) and vary the texture, while keeping the grain size distribution, slip system strength, slip hardening rates, strain rate and other factors constant while simulating uniaxial tensile deformation. 

\subsubsection{Simulating Uniaxial Tensile Deformation}
Due to the homogenization approach, mean field models of crystal plasticity cannot account for neighboring grain interactions and the intragranular heterogeneity in the micromechanical fields. Hence a full field elasto-viscoplastic fast Fourier transform (EVPFFT) crystal plasticity formulation (\cite{Lebensohn2012}) is chosen to simulate uniaxial tensile deformation in the generated microstructures. This model takes in the microstructure image discretized on a regular Fourier grid and the material properties such as the elastic stiffness tensor and slip systems for plastic loading. The output is the stress, strain-rate and orientation fields at each grid point. Since the EVPFFT model solves the constitutive equations at each grid point, the grain size should not affect simulations as long as the Fourier grid size is chosen such that the model converges. The boundary conditions on the strain are chosen such that the material transitions into the plastic regime. 

The  Voce hardening law \cite{voce1955practical} is used to model the evolution of the critically resolved shear stress (CRSS) of each slip system s, $\tau^s$, with accumulated shear strain as follows:
\begin{equation}
\tau^{s}(\Gamma) = \tau^{s}_{0} +(\tau^{s}_{1}+\theta^{s}_{1}\Gamma)\left(1-exp\Big(
-\Gamma\mid \frac{\theta^{s}_{o}}{\tau^{s}_{1}}\mid\Big)\right)
\end{equation}
The parameters $\tau_0$ and $\theta_0$ refer to the initial yield stress and the initial hardening rate. $(\tau_0 + \tau_1)$ is the back-extrapolated stress and $\theta_1$ is the asymptotic hardening rate. $\Gamma$ is the accumulated shear in the grains. The Voce hardening parameters were extracted as shown in table \ref{Voce} by fitting the VPSC simulated stress-strain curve to representative experimental stress-strain curves for FCC copper. \ref{AppendixA} covers the details of extracting these parameters.


\subsubsection{Problem Formulation: defining stress hotspots}
\begin{figure}[!hbt]
\centering
    \begin{subfigure}[t]{0.4\textwidth}
        \centering
        \includegraphics[width=\textwidth]{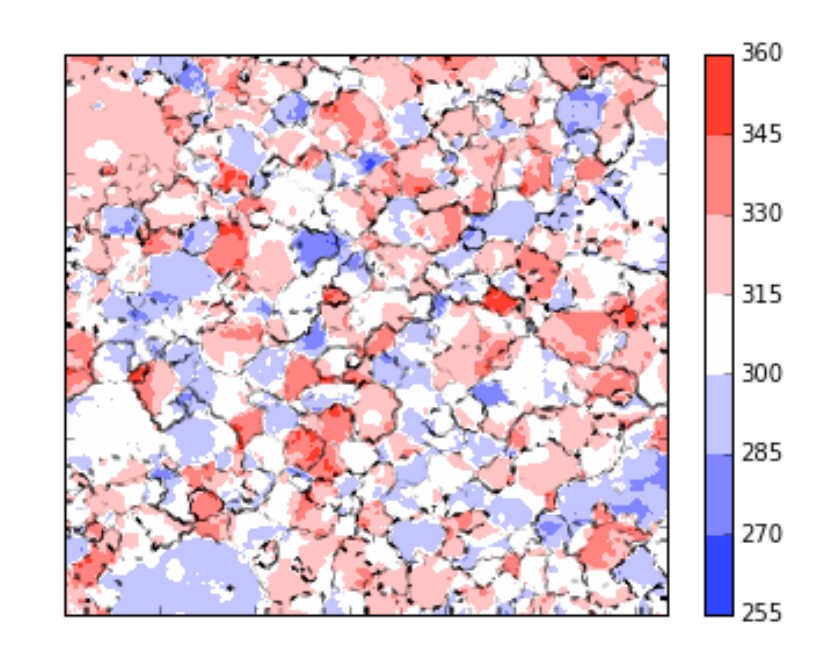}
        \caption{ }
        \label{fig:stressmap}
    \end{subfigure}
     \begin{subfigure}[t]{0.4\textwidth}
        \centering
        \includegraphics[width=\textwidth]{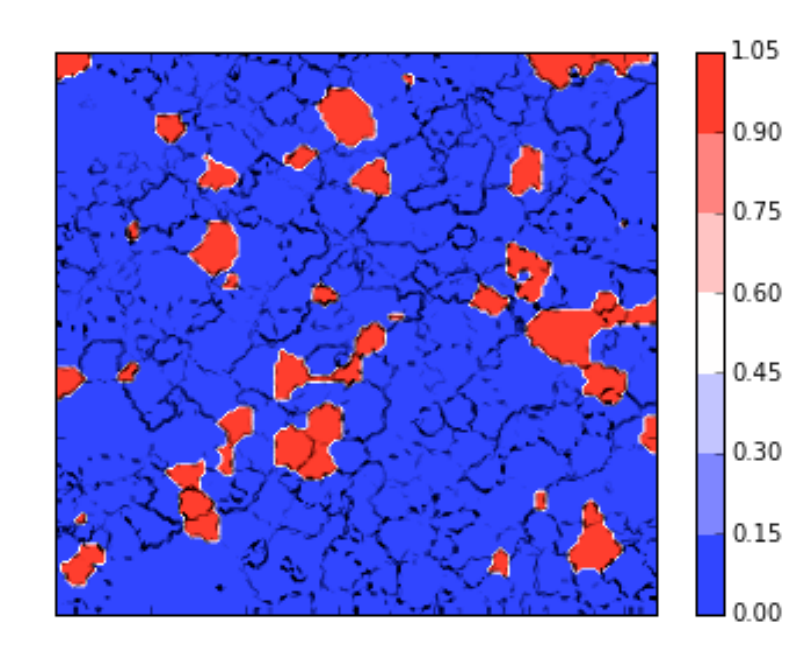}
        \caption{ }
        \label{fig:hotgrainmap}
    \end{subfigure}
    \caption{Cross sections of a microstructure showing (a) Von Mises Stress field and (b) Stress hotspots obtained by thresholding on grain averaged stress field. The red regions correspond to regions of high stress or stress hotspots. } 
    \label{fig:hotgrainpoint}
\end{figure}

EVPFFT simulations result in a voxel-wise output for the Von Mises (VM) stress field, as shown in Figure \ref{fig:stressmap}. It is observed that the regions of high stress generally form in clusters and intra-grain variations in stress values are small. Therefore, to minimize the impact of numerical artifacts and small-scale fluctuations, this field is averaged grain-wise to get the stress in each grain. The resultant stress distribution is thresholded using the peak over threshold method (\cite{Donegan2013}) to select the critical stress threshold (Figure \ref{fig:hotgrainmap}). The grains having VM stress above the critical threshold are designated as stress hotspots. The critical threshold grain averaged Von Mises stress value was found to lie between the $85^{th}$ and $95^{th}$ percentile. Hence the $90^{th}$ stress percentile was chosen as a cutoff throughout the dataset to keep the fraction of hotspots the same between all the microstructures. 

\subsubsection{Relation Between Stress Hotspot locations}\label{Relation Between Stress Hotspot locations}
A spatial autocorrelation analysis was performed to  determine if the location of a hot grain has a role in the formation of another hotspot, i.e. if the hot grains are related to each other.  Two-point statistics are calculated between hot grains and normal grains using the method described in \cite{Fast2011}. We observe a high autocorrelation inside the coherence length which represents the average size of the hotspots. Outside this length, we find that the probability of finding a hotspot is constant in all directions, which means that the hot grains are not correlated in space i.e. they are dispersed uniformly in the microstructure. Thus stress hotspots (grains) are independent of each other and standard statistical approaches can be used without having to worry about the shadowing effects of spatial correlation.

\subsubsection{Effect of Microstructure evolution on Stress hotspots}\label{Effect of Microstructure evolution}
As a material deforms, the grains tend to rotate towards the tensile axis, thus changing their orientation. \cite{Lebensohn2012} have used the EVPFFT formulation to study hotspots formed during uniaxial tensile deformation in polycrystalline FCC materials. It was observed that in materials where the "hard" and "soft" directions in the elastic and plastic regimes are different, the elastic hotspots become plastic cold-spots and vice-versa. 

\begin{figure}[!hbt]
\centering
    \includegraphics[width=0.8\textwidth]{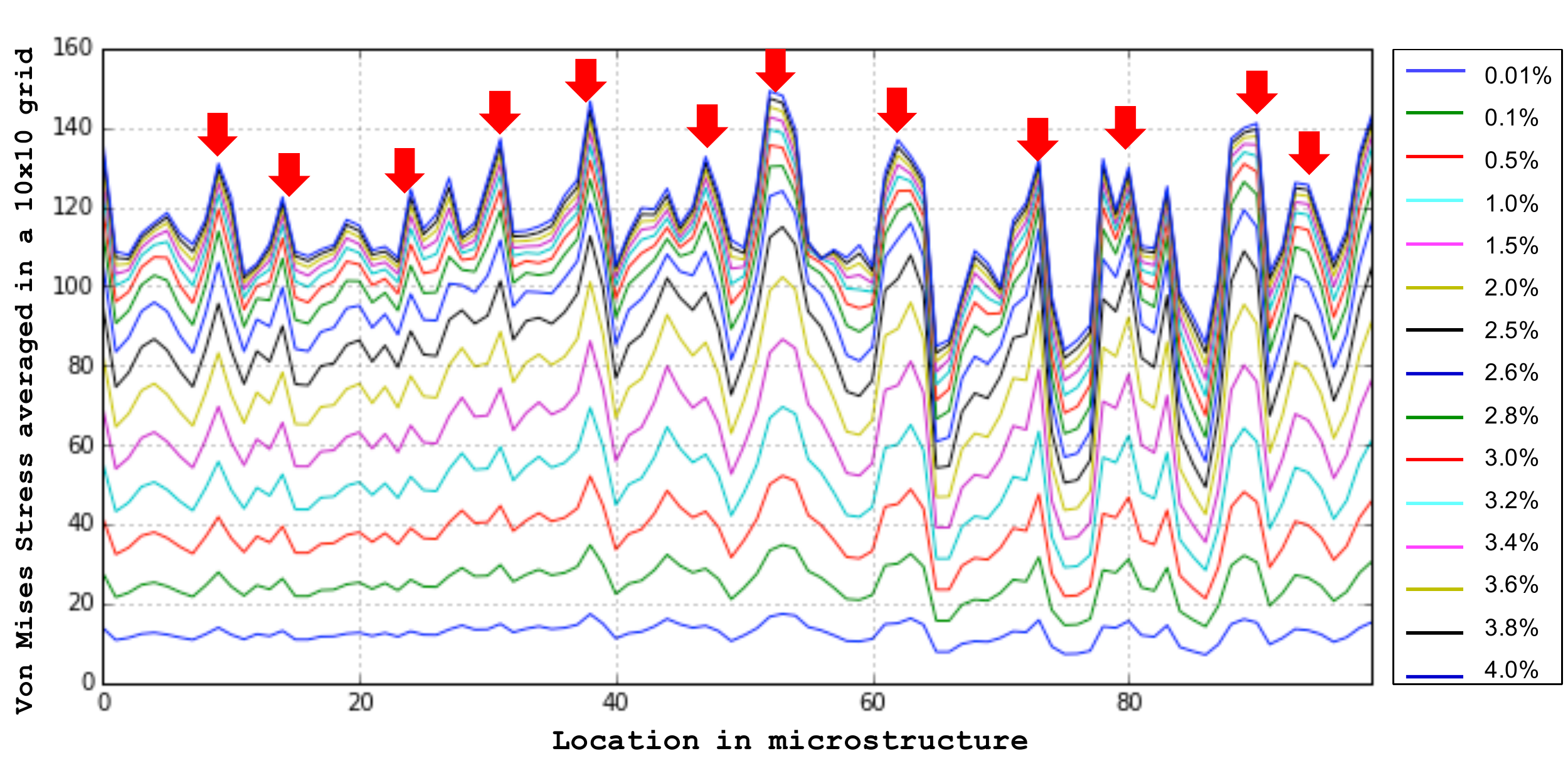}
    \caption{Evolution of Von Mises stress vs. location in a cross section of the microstructure at each deformation step. The bottom (blue) curve is at the lowest strain (0.01\%) and the top (blue) curve is at the maximum strain (4\%). The red arrows indicate how the location of stress hotspots is not changing}
    \label{HotspotEvolve}
\end{figure}

The constitutive model parameters for FCC materials represent a generic copper alloy.  For the single crystal elastic constants used in this work, the elastic anisotropy parameter given by equation \ref{A_copper} (\cite{zener1948elasticity}) is $A = 3.2$. For $A > 1$; $<100>$ and $<111>$ are soft and hard elastic directions respectively. This coincides with the plastic anisotropy of $\{111\}<110>$ slip (\cite{Lebensohn2012}), not taking the effect of strain rate into account. Hence, elastic hotspots should remain plastic hotspots (and cold spots remain cold) as the deformation proceeds.

\begin{equation}
A = (2\times C_{44})/(C_{11} - C_{12})
\label{A_copper}
\end{equation}
To verify this, the location of stress hotspots in the elastic and plastic regimes was studied. A 128x128 cross section of the microstructure was divided into 196 10x10 regions. The Von Mises stress field was averaged in each 10x10 region, and is plotted at each deformation step in Figure \ref{HotspotEvolve}. It was found that the stress hotspots are stationary in the crystal structures studied. That is, initial high stress regions remain high in stress as deformation proceeds; likewise, low stress regions retained low stress values throughout deformation.

\subsubsection{Effect of Texture and Microstructure}\label{Effect of Texture}
We conducted two experiments. First we compared the location of stress hotspots in microstructures having the same grain structure but different texture templates applied to it. In the second experiment, we compared the location of stress hotspots in microstructures having similar random textures, but stochastically different microstructures. All the microstructures have equiaxed grains and a lognormal grain size distribution with mean grain size of 2.7 microns. We found that changing either the texture or the grain structure has an impact on the macroscopic flow stress-strain response and the location of hotspots.

\subsection{Developing Microstructural Descriptors}
Feature engineering is the process of using knowledge of the pertaining field (in this case, deformation mechanics) to build features to be used by a machine learning algorithm (\cite{Domingos2012}). During material deformation, the loading condition and microstructure are the most important factors that determine stress distribution in a microstructure. For stress hotspot prediction, both crystallographic and geometric microstructural descriptors form the domain knowledge based feature set.  Crystallography and geometry based microstructural representations to be used as features during machine learning are developed in the next sections. \ref{Table of Acronyms} lists the acronyms and descriptions of the features used in this work.

\subsubsection{Crystallographic Descriptors}
The shear strain rate on a material point depends on the tensor dot product between the Schmid tensor and the deviatoric stress tensor. These tensor quantities are determined by the grain orientation, which is captured by the Euler angles (\cite{KocksTomeWenk, Piehler2009}). However, due to the non-linearity of Euler space, there can be multiple Euler angles representing the same orientation. Interpreting the results from Euler angle ranges associated with hotspots is difficult even after reducing the Euler angle space to the fundamental zone. The complex, non-linear, trigonometric relationship between the tensor quantities cannot be captured directly by using Euler angles as features, and hence we develop the following features to represent the grain crystallographic properties.

\begin{figure*}[!htb]
\centering
    \begin{subfigure}[t]{0.33\textwidth}
        \centering
        \includegraphics[width=\textwidth]{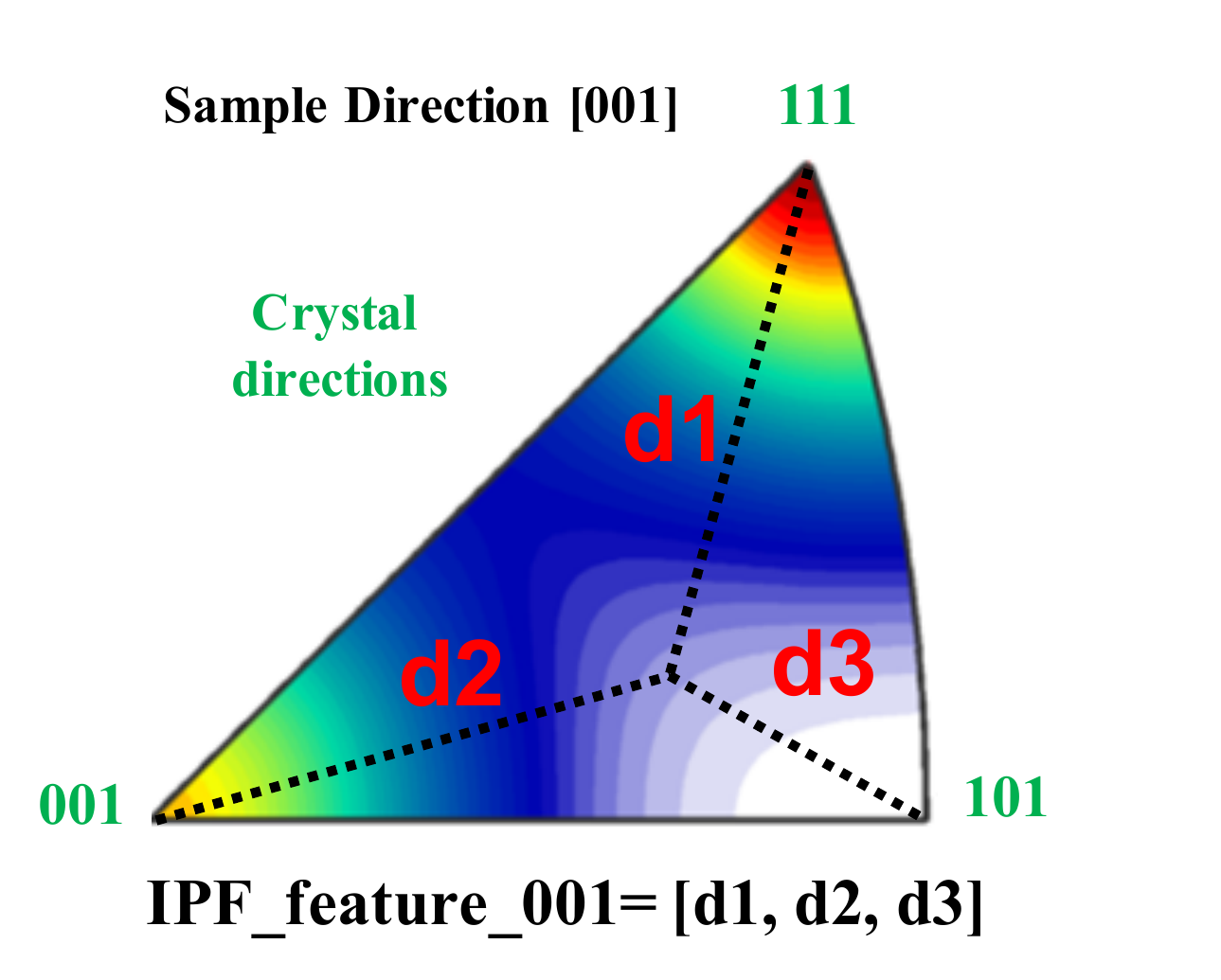}
        \caption{}
        \label{fig:IPFschematic_1}
    \end{subfigure}
    \hspace{\fill}
    \begin{subfigure}[t]{0.25\textwidth}
        \centering
        \includegraphics[width=\textwidth]{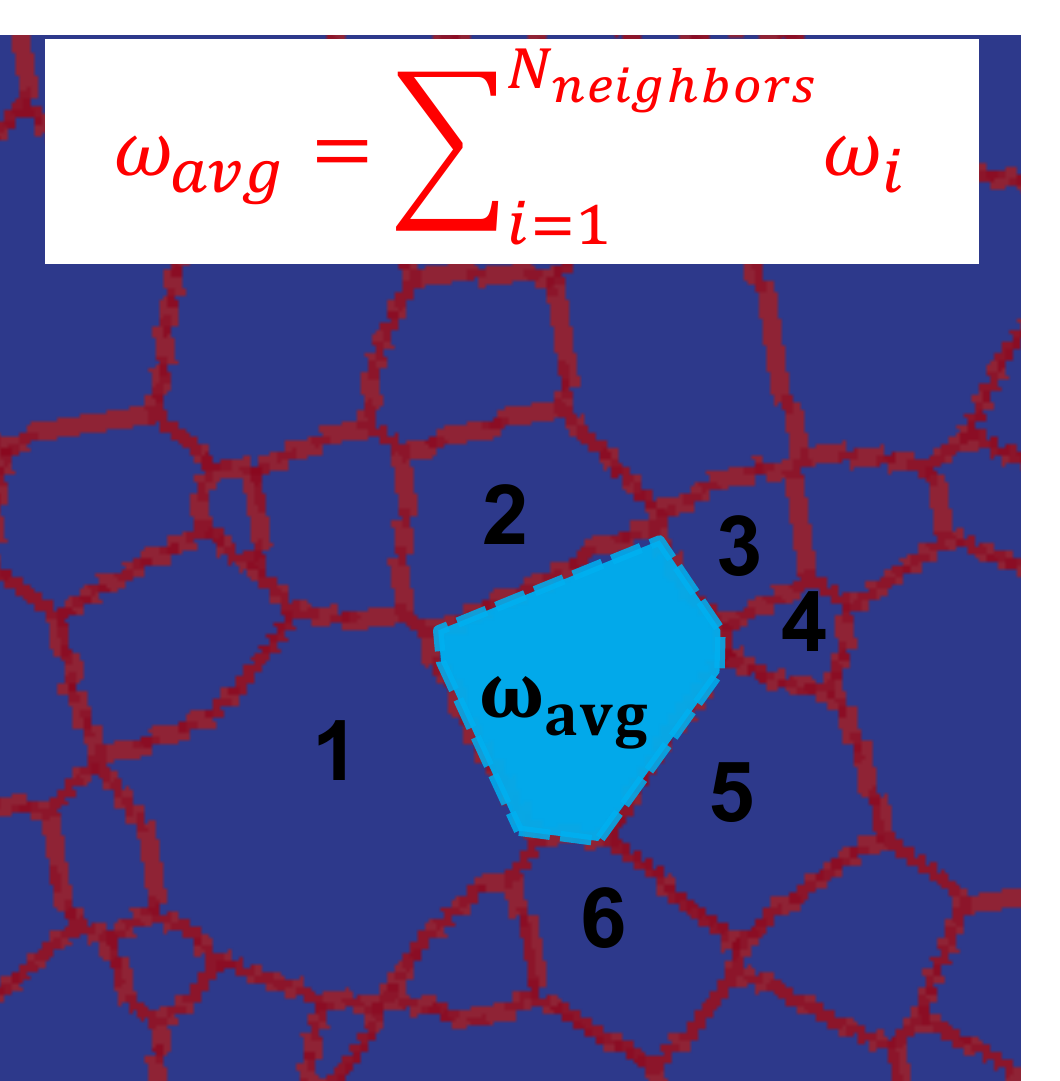}
        \caption{}
        \label{fig:misorientation}
    \end{subfigure}
    \begin{subfigure}[b]{0.4\textwidth}
        \centering
        \includegraphics[width=\textwidth]{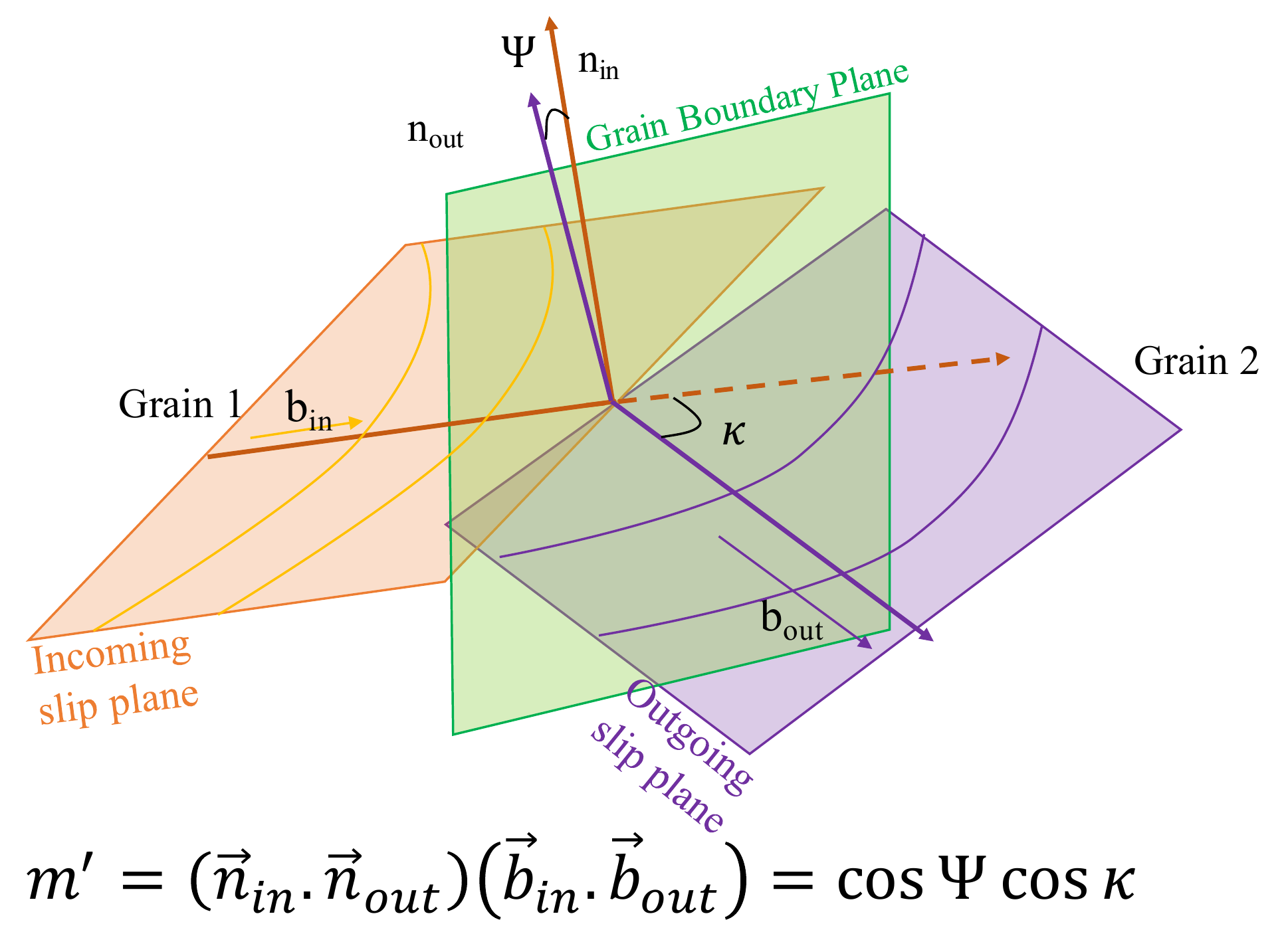}
        \caption{}
        \label{fig:stm}
    \end{subfigure}
    \caption{Crystallographic descriptors used as features during machine learning. (a) Schematic of an Inverse pole figure for a cubic crystal. The distance of the [001] sample direction from the three crystal directions [001] (d2), [101] (d3) and [111] (d1) in the standard stereographic triangle are used as a feature. (b) Schematic of average misorientation $\omega_{avg}$ for the reference (light blue) grain with respect to it's N contiguous neighbors (numbered 1 to 6) in a 2-D microstructure. $\omega_i$ is the misorientation between the reference grain and its $i^{th}$ neighbor grain. (c) Slip Transmission Factor (m-prime, $m'$) is a function of the dot product of the slip plane normals (incoming normal $n_{in}$ and outgoing normal $n_{out}$), and the dot product of slip directions  (incoming $d_{in}$ and outgoing $d_{out}$). Adapted from \cite{mprime}}
    \label{fig:123}
\end{figure*}

\textbf{Distance from Inverse pole figure corners: }
A microstructural descriptor based on the inverse pole figure (IPF) space is proposed to capture the relation between the tensile axis direction and the grain soft and hard orientations resulting from the anisotropy in elastic modulus. The loading direction is projected in the inverse pole figure space in the fundamental zone. This ensures that every orientation is projected into the same stereographic triangle. The distance of each projected point from the three corners of the inverse pole figure describes the orientation of the loading direction w.r.t. the 3 crystal directions. Since this space is fixed, the Euclidean distance of the projected point from the 3 corners of the inverse pole figure is invariant and is used as a microstructural descriptor. This distance is an indicator of how close the tensile axis is to the crystal directions in the inverse pole figure. Figure \ref{fig:IPFschematic_1} shows a schematic of the three distances in cubic materials. 


\textbf{Misorientation: }
\cite{Aust1954} show that during aluminum bicrystal deformation, the yield stress and the rate of work hardening increases with the orientation difference between the crystals. Some misorientations correspond to "special" grain boundaries which could lead to enhanced properties with respect to corrosion, impurity segregation, cracking, coarsening, diffusion and other properties affected by grain boundary properties (\cite{Lehockey19983069}). Hence, the misorientations of grains with their $1^{st}$, $2^{nd}$ and $3^{rd}$ neighbors were calculated, and features like minimum, maximum and mean misorientation for a grain were calculated from these lists. Figure \ref{fig:misorientation} shows a grain, it's contiguous neighbors and the average misorientation calculation in a 2-D microstructure.

\textbf{Slip Transmission Metrics: } 
A geometric compatibility factor to measure the ease of slip transmission between two grains is the mprime factor from \cite{Luster1995}, calculated as a dot product between the slip plane normals and slip directions across a grain boundary as seen in figure \ref{fig:stm}. The m-prime factor lies in the range (0,1); a value of 0 indicates incompatible deformation across the grain boundary and a value of 1 indicates co-planarity of slip systems between two grains. \cite{Bieler2014} show that this factor is related to the local misorientation between the slip systems in different grains and can help in predicting the slip transmission across different grains. This metric is calculated using a filter in Dream.3D for FCC materials.

\textbf{Schmid Factor: }
The Schmid factor is a measure of the optimal orientation of a slip system for deformation in a single crystal, and can be extended to polycrystalline materials (\cite{Piehler2009}).

\textbf{Other Factors}
The \textit{Taylor factor} (\cite{taylor1938plastic}) consists of information about the slip occurring in each grain and the equivalent Von Mises stress and is calculated during deformation. Since our goal is to correlate only the initial microstructural descriptors to stress hotspots, the \textit{Taylor factor} is not included in the feature set.  Factors like \textit{stress triaxiality, slip system activities, principal stress directions} and \textit{Lode angle parameters}, are important in determining the stress state in a grain, but are dropped from the feature space for similar reasons.

\subsubsection{Geometrical Descriptors: Grains and their neighborhood}
Material failure is highly dependent on microstructure. For example, cleavage fracture in mild steels has been shown to have a grain size dependence (\cite{Smith1968}), whereas during ductile fracture under uniaxial tensile deformation, stress hotspots tend to form near microstructural features and usually form in textures corresponding to maxima in Taylor factor (\cite{Rollett2010a}). In this section, we  develop a few microstructural descriptors describing the grain geometry which are used in this work.

\textbf{Shape averaged distance from special points: }Due to the three-dimensional polycrystalline grain structure, the grain boundary network consists of grain boundaries where two grains meet; \textit{triple lines}, where 3 grains and 3 boundaries meet; and \textit{quadruple points}, where 4 grains, 6 boundaries and 4 triple lines meet (\cite{Smith1948}). The grains closer to these special points have a neighborhood with a higher local heterogeneity. The triple junctions, depending on their crystallography, can impede or permit inter-granular damage through them (\cite{JohnsonSchuh2013}), and hence are an important microstructural descriptor.  \cite{Rollett2010a} show that stress hotspots tend to form in grains closer to these special points . Hence we use the average distance of each grain to the closest grain boundaries, triple lines, and quadruple points as geometric features. A Dream.3D filter is used to calculate the Euclidean distances for each voxel, which are then averaged per grain to form a grain level feature. 

\textbf{Grain shape parameters: }
\cite{Rathmayr2013} show that in severely plastically deformed (SPD) materials, such as those produced by high pressure torsion (HPT), the ductility is influenced by the grain aspect ratio. In alpha-beta Titanium alloys, the grain size, aspect ratios and grain size distribution have been shown to be related to the flow stress (\cite{Rhaipu2002}). The Hall-Petch and inverse Hall-Petch relationships describe the relation between yield stress and grain size in a material (\cite{counts2008predicting}). In this work, we have used equivalent spherical grain diameter and grain aspect ratio from a best fit ellipsoid as features. They were calculated using a Dream.3D filter. The grains can also be characterized by the number of contiguous neighboring grains, which is used as a feature and calculated via Dream.3D.

\subsection{Machine Learning Methods}
The EVPFFT simulations and the derived microstructural descriptors form a sample on which machine learning models can be built and evaluated. The microstructural descriptors are used as features to train machine learning algorithms for predicting stress hotspots. The contribution of each microstructural feature in predicting hotspot formation is studied to understand it's impact. The machine learning methods used in this work are described in this section.


Machine learning (ML) techniques are well established and have been applied to different fields such as advanced manufacturing, financial services, health care, marketing and sales, robotics and transportation (\cite{LeCun2015, Mangal2017a}). Machine learning is a statistical framework that automates analytical model fitting for data analysis such as finding structure in data (clustering) and making data-driven predictions or decisions. Predicting stress hotspots is a binary classification problem. Given a grain in a microstructure, we want to predict if a stress hotspot forms there. A feature vector $\mathbf{X}$ whose components are derived from the microstructural descriptors is constructed. ML methods can be used to extract insights and correlations between the elements of $\mathbf{X}$ and the micromechanical outcome. 

\subsubsection{Model Performance Metrics}
Once a model is trained, it's performance can be evaluated by looking at the classification accuracy. However, for stress hotspot classification, predicting all the grains as normal (non-hotspots) will still result in a $90\%$ classification accuracy as only $10  \%$ of the grains are hotspots.  A better representation of the classification accuracy is the two-dimensional confusion matrix, indexed in one dimension by the true class and in the other by the predicted class. A correctly predicted hot grain is true positive; a misclassified hot grain is a false negative. The normal grains predicted as hot are false positives, and the normal grains predicted normal are true negatives. We want the classifier to correctly predict all hotspots, maximizing the true positives while minimizing the false positives. The receiving operator characteristic curve is the plot of false positives vs. true positives. The area under the receiving operator characteristic curve (AUC) is a good evaluation metric for such unbalanced datasets (\cite{auc}). If the classifier is no better than random guessing, the true positive rate will increase linearly with the false positive rate and the area under the receiving operating characteristic curve will be around 0.50. A good classifier has a high true positives rate and a low false positive rate, and the $AUC \sim 1$.

\subsubsection{Estimation of Model Generalization Error}
A data-driven model should achieve good performance on the training data as well as generalize well on unseen (test) data. We divide our dataset into training and test (holdout) sets. During training, the model parameters are optimized using K-fold cross validation (CV) (\cite{Kohavi2016}). In this technique, the training sample is randomly partitioned into k subsamples. Then (k-1) subsamples are used to train the model, which is validated on the $k^{th}$ subsample. This process is repeated k times (the folds), such that each fold is used exactly once for cross validation. The k results are then averaged to get the validation estimation. To offset the effect of unbalanced data, stratified k-fold cross validation is common (\cite{Parker2007}), where the k-folds are selected such that they have approximately equal proportion of the class labels. In this work, the grain-wise hotspots are designated based on the crystal plasticity simulations and might be correlated within a microstructure as they come from the same crystal plasticity simulation. Therefore, to assess the generalization error during validation and test times, we perform stratified sampling. The validation folds are created by selecting the grains from a randomly chosen microstructure per texture class that is absent in the training data. This overcomes the optimistic bias in generalization error from having correlated data from a single simulation between training and validation. Once the model hyper-parameters are optimized, a second K-fold cross validation is run by using the entire training dataset to construct the models and creating validation folds from the microstructures in the holdout (test) dataset. This helps in getting an estimate of the generalization error on unseen data. This technique is also known as nested cross-validation (\cite{Varma2006}).

The generalization error can be understood by dividing it into bias and variance errors. If the learning algorithm is too simple, it introduces a bias in the predictions, whereas a very complex algorithm can overfit and learn the noise in the dataset, leading to a higher variance in the predictions. Learning curves are diagnostic curves that can help in understanding the trade-off between bias error and variance error, thus helping the models to generalize beyond the training dataset. They are a plot of the model evaluation metric on the training and cross validation datasets as a function of the training dataset size. 

The model performance generally increases with the training dataset size. If there is a gap between training and validation error/ performance, the model is suffering from high variance and collecting more data will help. On the other hand, the model suffers from bias when training and testing errors converge and are high, and a more complex algorithm or more features are needed. These curves can be used to determine the size of training dataset required.  

\subsubsection{Random Forest Model} 
\begin{figure}[!htb]
\centering
	\includegraphics[width=0.4\linewidth]{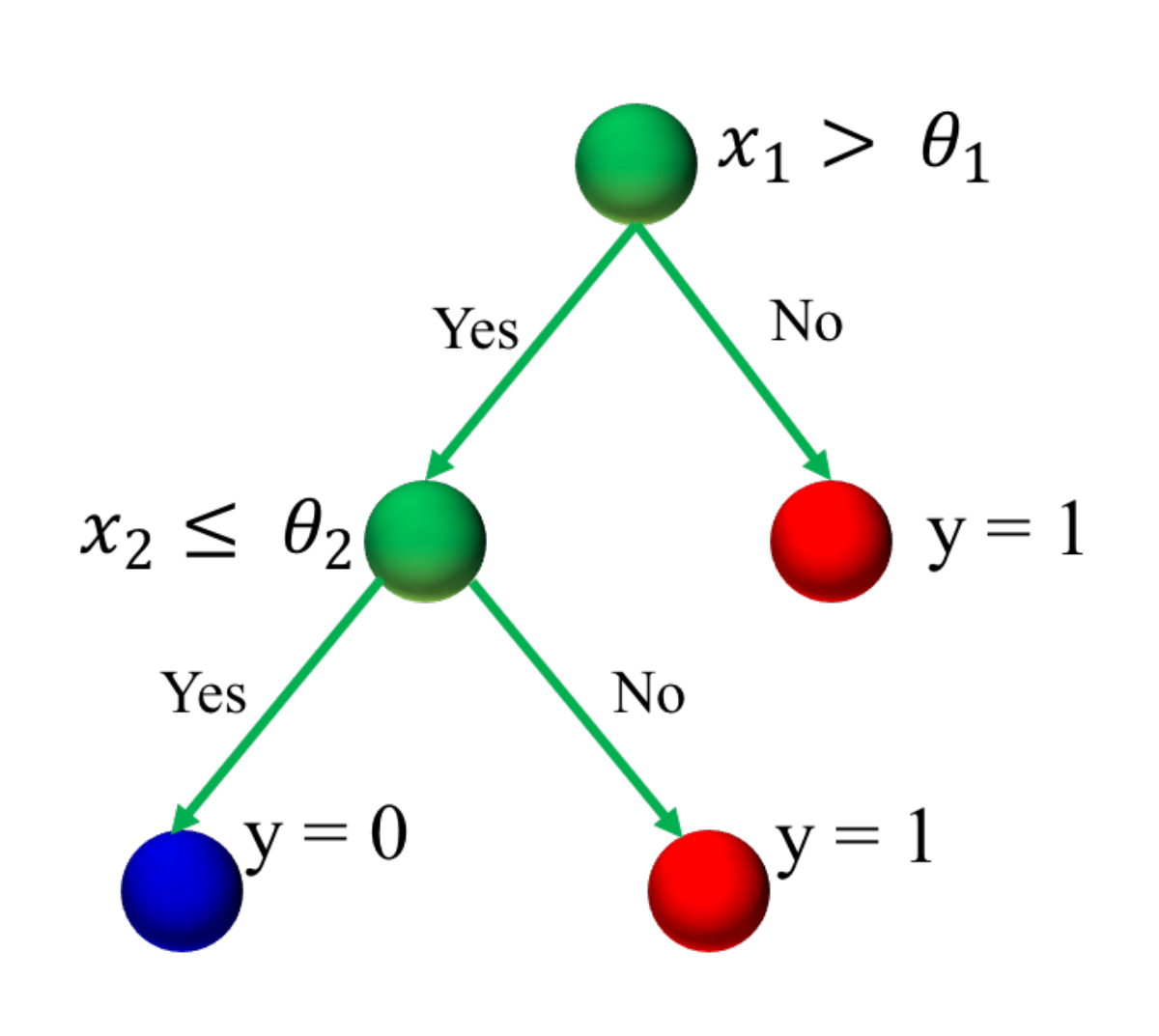}
	\caption{Schematic of a decision tree model built on two features $x_1, x_2 \in \mathbf{X}$ with output class labels $y\in \{0,1\}$. The split at the nodes is decided on the value of $x_1, x_2$ which maximizes the information gain (\cite{Quinlan1986, quinlan2014c4}) }
	\label{fig:DecisionTree}
\end{figure}

Machine learning methods are often viewed as black box approaches linking inputs to outputs using a complex set of functions. The main goals of this work is to understand the microstructural attributes causing hotspot formation. Hence we choose a non-black-box machine learning approach: a random forest (RF) algorithm which is built  on decision trees (\cite{Breiman2001}). A decision tree is like a flowchart, with every node containing a test on the features and the branches leading to the leaf node which represents the output class label (Figure \ref{fig:DecisionTree}). The RF algorithm utilizes ensembling to bring together a number of weak decision trees. This averages out the bias, reduces variance and avoids overfitting issues common in simple decision trees (\cite{dietterich2000ensemble}). Each tree is built on a random subset of the training data, using a random subset of the feature set thus bringing stochasticity while training the model which helps in avoiding overfitting the training dataset. The trees are grown greedily by choosing the split on the variable that minimizes the Gini impurity. The output from each of the decision trees is then voted to get a final prediction. Note that we tried a number of other tree based models like XGBoost (\cite{Chen:2016:XST:2939672.2939785}) and Gradient Boosted trees which did not result in an improvement in the model AUC. 

RF models are very fast and easy to fit, can handle both numerical and categorical features and deal with missing or unbalanced data efficiently. The number of decision trees, the number of features, and the depth of the decision tree are model hyper-parameters. The hyper parameters are chosen using a random grid search by comparing the cross validation performance. The optimized values are max\_depth=8 and num\_estimators=1200. The RF algorithm was implemented in Python using the scikit-learn implementation (\cite{pedregosa2011scikit}). In this implementation, the predicted class is the one with the highest mean probability estimate across the trees in the random forest. 

Our dataset was constructed for a set of representative textures shown in Figure \ref{fig:FCCtex}. For each texture kind, stochastically different datasets are created via multiple microstructure instantiations. The location of stress hotspots is affected by texture, geometry and the constitutive parameters. For materials consisting of equiaxed microstructures under uniaxial tension, we compare two machine learning frameworks to capture the variation caused by a texture kind:
\begin{itemize}
    \item \textbf{Partition models}: a different RF model is trained for each texture kind and the AUC score is reported using k-fold cross validation for each model (average of validation performance on each microstructure in a texture kind).
    \item \textbf{Mixed-model}: a single RF model is trained on all the microstructures with different textures, and the AUC score is reported using k-fold cross validation (average of validation performance on 2 randomly chosen microstructures from each texture kind).
\end{itemize}

\subsubsection{Feature Importance Metrics}
Random forest models have an embedded metric known as the permutation accuracy importance (PAI) (\cite{Breiman1996, Breiman2001}), which can be used to get feature importance scores to understand the contribution of each feature in predicting stress hotspots. However, this metric is prone to correlation bias due to preferential selection of correlated features during the tree building process (\cite{Strobl2008}). It is essential to choose the right feature importance metrics to get data driven insights in order to avoid incorrect conclusions (\cite{mangal2018comparative}). Hence, to gain data driven insights, we compute feature importances using state of the art FeaLect method (\cite{Zare2013}). It is a robust feature selection method that computes the feature importance using LASSO (L1) regularization (\cite{Tibshirani1996}). We first oversample the dataset to balance the population of the two classes. The oversampled dataset was bootstrapped (i.e. drawing random samples with replacement) 100 times. In each random subset, various linear models are fitted using the lars method (\cite{Efron2004}), maintaining the regularization strength such that only 10 features are selected by LASSO. Features are scored in each model depending on their tendency to be selected by LASSO in each model. Finally, these relevance-orderings are averaged to give the feature importance on an absolute scale. We used the R implementation of FeaLect to compute our results (\cite{Zare2015}).


\section{Results and Discussion }

\begin{figure}[!btp]
  \centering
  \begin{tabular}{p{3in}p{4in}}
  \multirow{2}{*}[\dimexpr1.5in-0.01ex\relax]
  {
  \begin{subfigure}[t]{0.6\textwidth}
		\centering
	    	\includegraphics[width=0.65\textwidth]{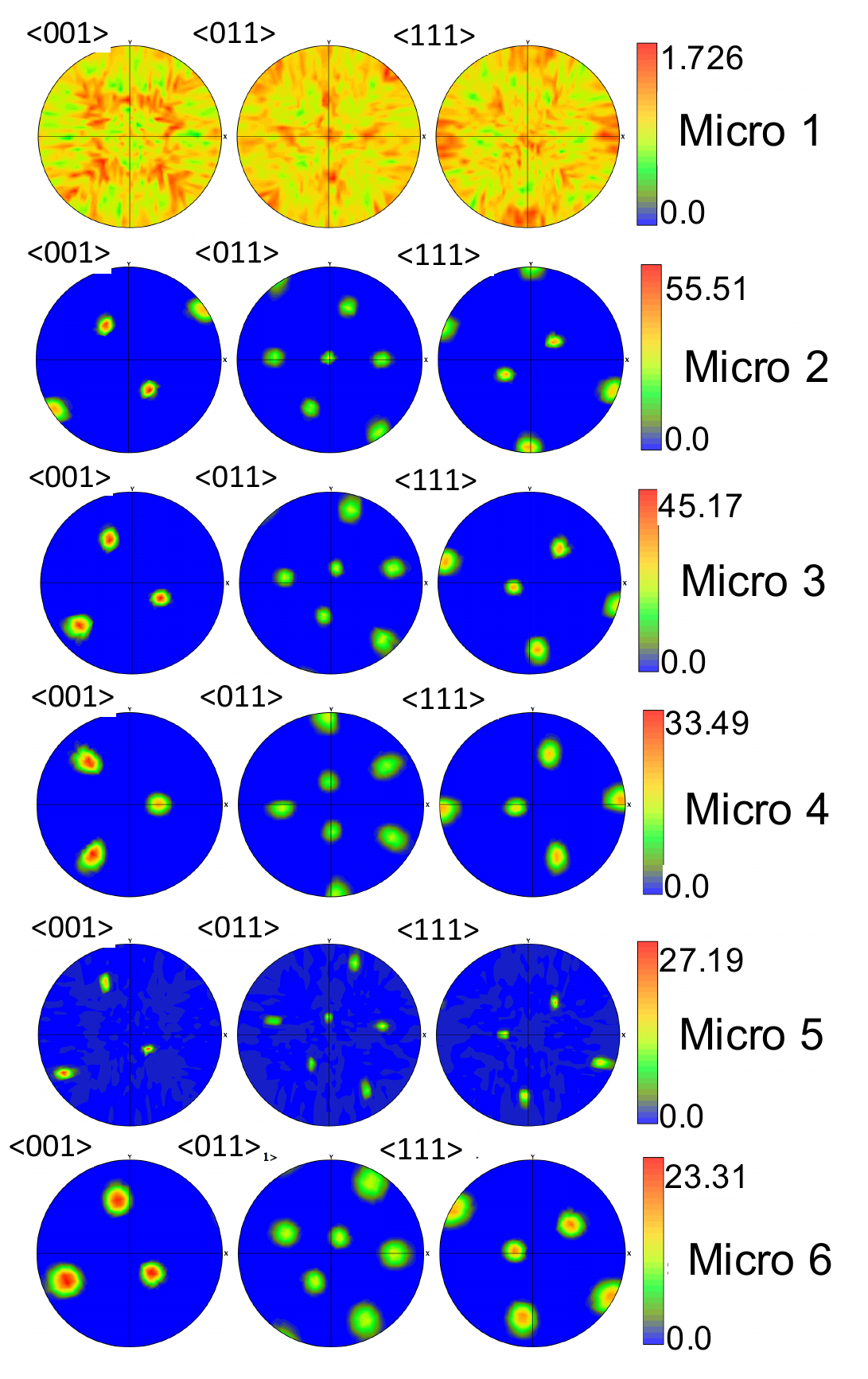}
	     	 \caption{ }
	    	  \label{fig:FCCtex}
	\end{subfigure}
	} &
	\begin{subfigure}[t]{0.4\textwidth}
	        \centering
		\includegraphics[width=\textwidth]{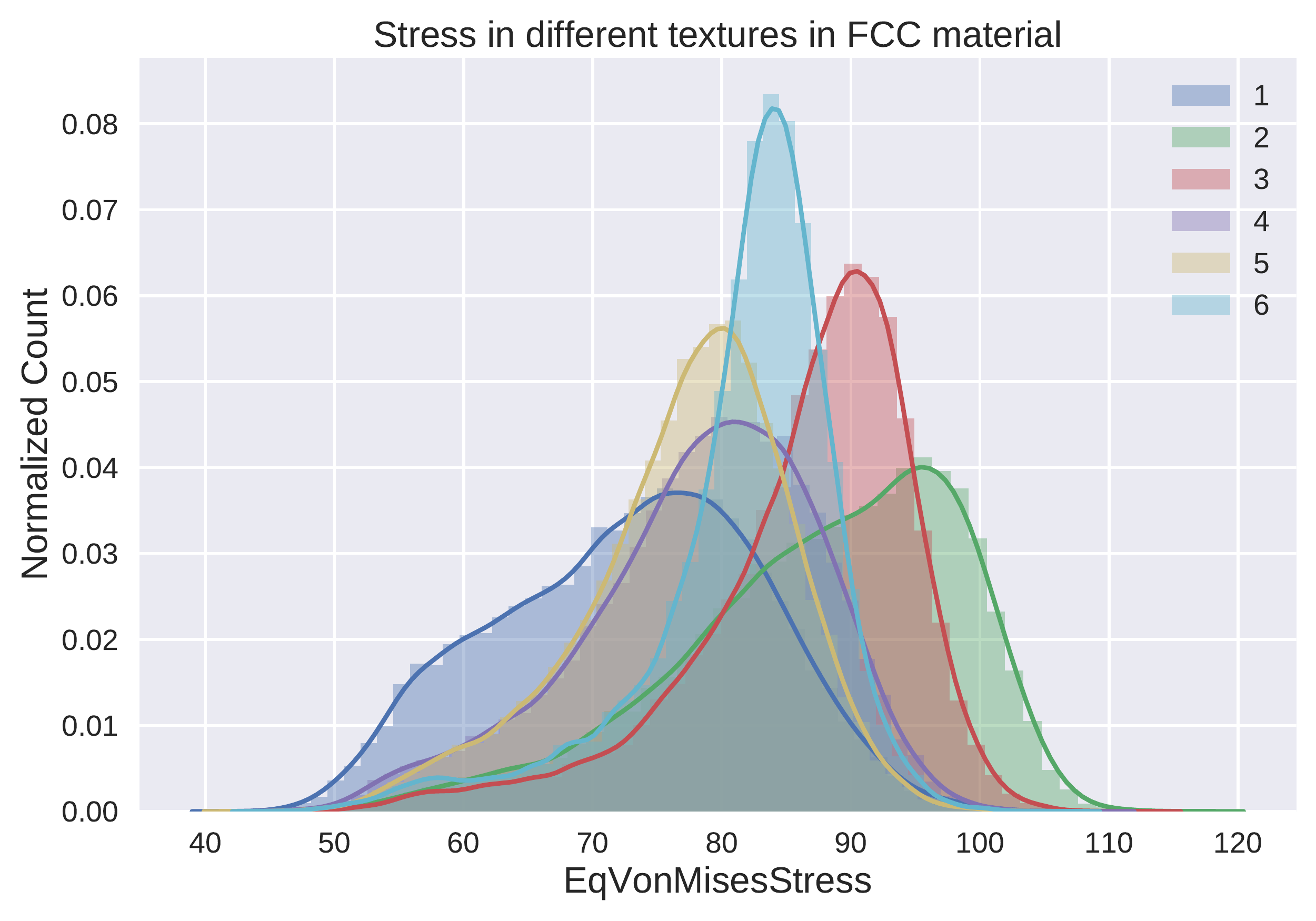}
   		 \caption{ }
    		\label{fig:FccDatasetStress}
    	\end{subfigure}
	\\
  & \begin{subfigure}[t]{0.4\textwidth}
    \includegraphics[width=\textwidth]{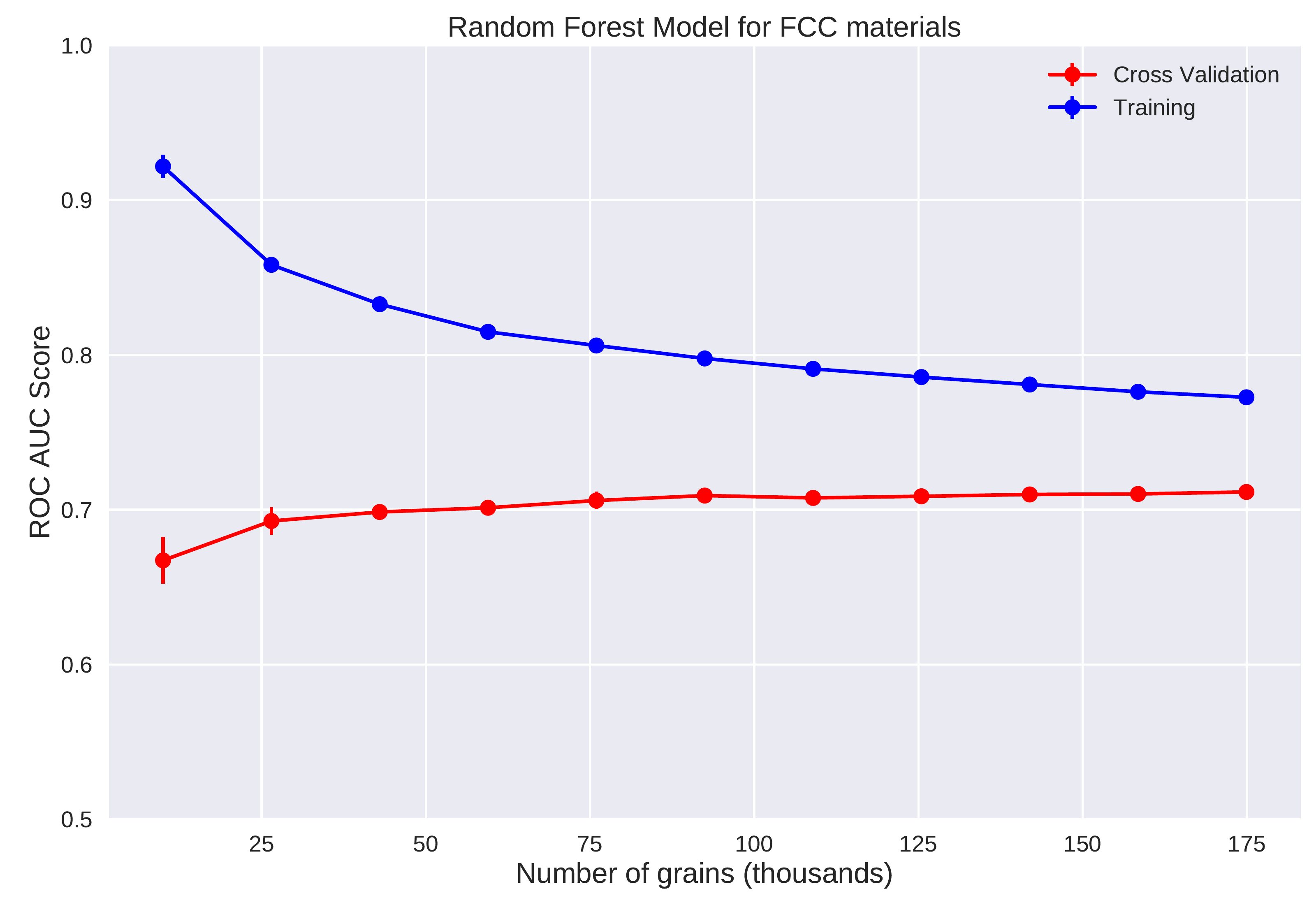}
        \caption{ }
        \label{FCC learning curve}
    \end{subfigure}
  \end{tabular}
  \caption{(a) Representative pole figures for six different FCC textures. (b) Histograms of grain averaged stress in the different texture classes of and (c) Learning curve for random forest model built on all textures classes}
\end{figure}

\begin{table}[t]
\small
\begin{center}
\caption{Cross validation AUCs for FCC materials for mixed and partition models}
\label{AUC_fcc}
\begin{tabular}{*{3}{c}}
\toprule \textbf{Texture kind} & \textbf{Partition Model AUC (\%)} & \textbf{Mixed Model AUC (\%)} \\ \midrule
1&$71.27\pm1.25$&72.07\\
2&$73.91\pm3.69$&73.10\\
3&$66.13\pm2.74$&67.83\\
4&$73.61\pm6.90$&79.84\\
5&$74.01\pm3.25$&76.31\\
6&$72.65\pm3.07$&75.01\\
\midrule
All&$71.93\pm 3.88$ &$74.03\pm3.72$\\
\bottomrule
\end{tabular}
\end{center}
\end{table}

\begin{table}[!htbp]
\small
\begin{center}
\caption{Pearson Correlation Coefficients between features and Stress hotspots for FCC materials. }
\label{PearsonFCC}
\begin{tabular}{*{3}{c}}
\toprule \textbf{Feature} & \textbf{Correlation Coefficient} & \textbf{p-value}  \\ \midrule
001\_IPF\_0 (Distance between loading direction and $<100>$) &0.070&0.0\\
001\_IPF\_1 (Distance between loading direction and $<110>$)&-0.144&0.0\\
001\_IPF\_2 (Distance between loading direction and $<111>$)&-0.169&0.0\\
\midrule
Schmid Factor & -0.299&0.0\\
Average Slip Transmission Factor $m'$&0.038&0.0\\
Average Misorientation angle & -0.556&0.0\\
\midrule
Distance to nearest grain boundary &-0.004&0.13\\
Distance to nearest triple junction &-0.002&0.49\\
Distance to nearest quadruple point &-0.002&0.312\\
Grain Size (NumCells) & -0.003 & 0.275\\
FeatureBoundaryElementFrac & 0.004 & 0.16\\
\bottomrule
\end{tabular}
\end{center}
\end{table}

The FCC dataset consists of six different textures as shown in Figure \ref{fig:FCCtex}. Figure \ref{fig:FccDatasetStress} shows the grain averaged stress distribution in each texture class. The distributions are all left-tailed; both the magnitude and sharpness of the stress peaks vary with texture kind. From table \ref{AUC_fcc}, it is seen that a single model (Mixed-Model) built on all textures performs better at predicting hotspots than the set of Partition models built using the same set of features on each texture class separately. Not only is the Mixed Model average AUC higher than the average AUC for the Partition Models, but the Mixed Model also yields AUCs equal to or higher than the Partition Models for each individual texture class. The Mixed Model average AUC = $74.03 \pm 3.72 \%$, indicating hotspot prediction well above random chance. This signals the existence of overarching rules causing stress hotspots, independent of the individual textures.
 
From the learning curve for the mixed model shown in Figure \ref{FCC learning curve}, since there is a gap between the training and validation scores, we can conclude that the model is suffering from bias, and the model performance can be improved by either increasing the feature space, or by changing the model algorithm. It is difficult to argue whether we have exhausted the space of important crystallographic features, but not many geometric features have been utilized in these models. It is possible that since the model suffers from bias, the \textit{missing features} come from long range geometric features which are not captured in the current feature set.

\begin{figure}[!htbp]
    \centering
    \includegraphics[width=\linewidth, keepaspectratio]{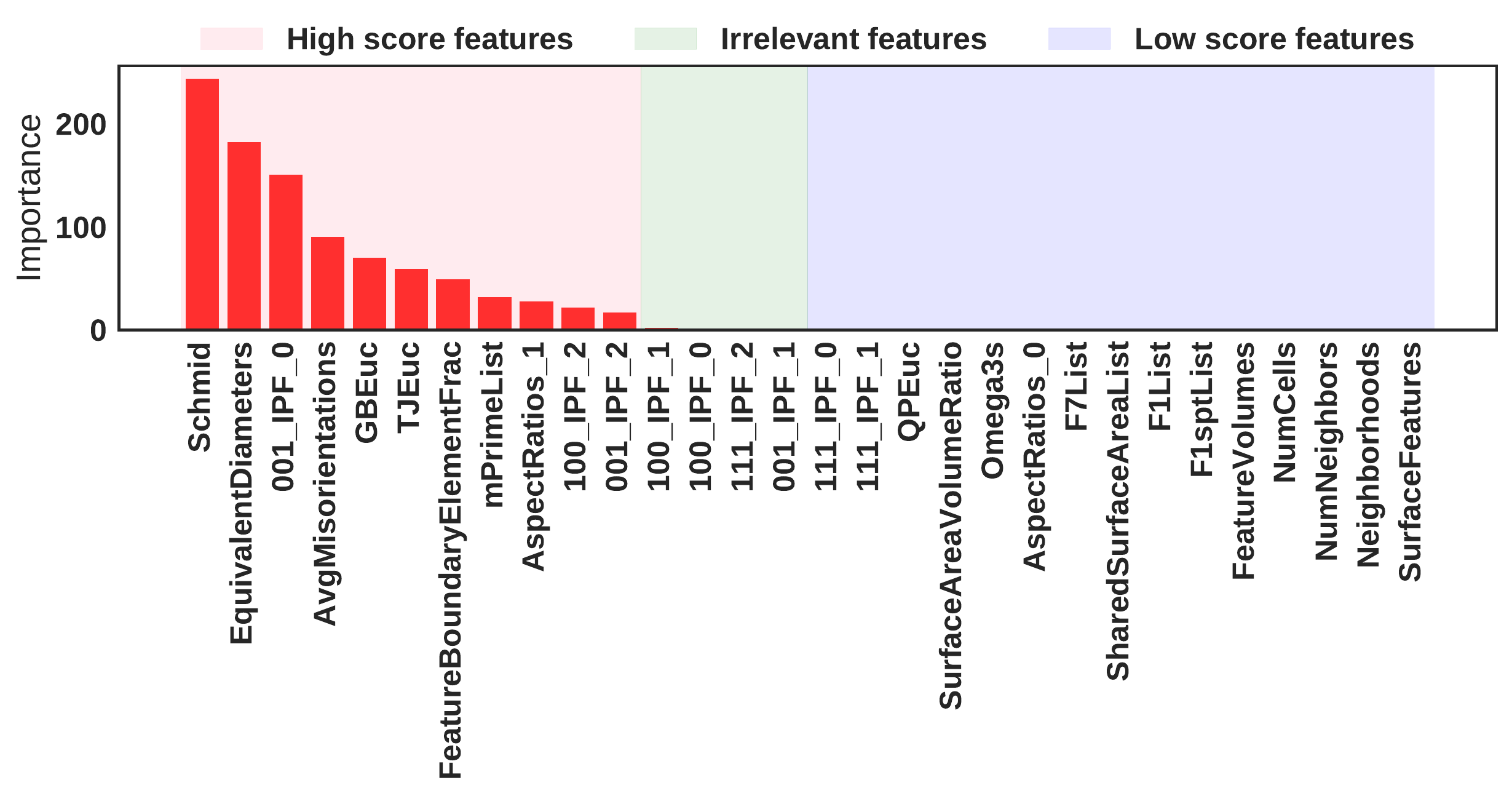}
    \caption{Variable importance in FCC materials: Both texture and geometry derived microstructural features are selected by FeaLect algorithm. The importance scale is arbitrary and set by FeaLect.}
    \label{fig:LassoFCC}
\end{figure}
Figure \ref{fig:LassoFCC} shows the feature importances for the mixed microstructure model calculated using the FeaLect method. The FeaLect plot reveals the set of most important (high score) informative features, which have minimum correlation between them. It also tells us the irrelevant features that might be responsible for overfitting in a linear model like LASSO (\cite{Zare2013}). The third set of features is the low score features which are least informative. However, for the random forest models, we found that removing the low score and irrelevant features do not significantly improve the model performance.

The Schmid factor of the $<110>\{111\}$ slip system is the most important feature. We calculated the Pearson correlation (\cite{Cohen2009}) between the important features and stress hotspots (table \ref{PearsonFCC}), and found that hotspots tend to form in grains with lower Schmid factor for FCC materials. 

Another important crystallographic feature selected by FeaLect is (Figure \ref{fig:LassoFCC}) the distance of the loading direction $[001]$ from the Inverse pole figure corner representing the $<100>$ crystal direction ($001\_IPF\_0$). The elastic modulus for FCC materials is anisotropic; it is highest along $<111>$ and lowest along $<100>$ crystal directions. The distance from the 3 corners of the [001] inverse pole figure projection gives an idea about the modulus of hot grains.  From table \ref{PearsonFCC} we can see that the individual Pearson correlations between the hotspots and these features are very small. We see a positive correlation for the distance of the loading direction from the $<100>$ corner ($001\_IPF\_0$) and negative for the others ($001\_IPF\_1$ and $001\_IPF\_2$). Thus the hot grains have a loading direction closer to the $<111>$ and $<110>$ corners as compared to the $<100>$ corner, i.e. hot grains have a higher elastic modulus.

The average misorientation and the slip transmission metric (m-prime) between a grain and it's nearest neighbors are the other selected texture derived features. From table \ref{PearsonFCC}, we see that hot grains have lower average misorientation and a higher slip transmission metric m-prime, i.e. they tend to be grains whose slip systems are more compatible with their neighbors in agreement with \cite{Plasticity2017}. We take this result with a grain of salt because dislocation movement occurs by a combination of active slip systems and geometrically favorable slip transmission.

Among the geometry derived features, we see that grain size, shape averaged distance from triple junctions and grain boundaries, grain aspect ratio and the fraction of grain lying on the periodic boundary during simulation are important. From table \ref{PearsonFCC}, we see that hotspots lie closer to grain boundaries, triple junctions and quadruple points i.e. they form in smaller grains. This result is in agreement with \cite{Rollett2010a} where stress hot spots were found to lie closer to microstructural features. 

It has been shown that the stress field calculated by EVPFFT can spike at the periodic microstructure boundaries, because voxels at the surface are different from the interior (\cite{Anglin2014}). The fraction of a grain lying on the periodic boundary is an important feature in deciding stress hotspots, and captures this simulation artifact. 

Taken together, these results demonstrate that both crystallographic and geometric features contribute to the formation of stress hotspots. The power of machine learning and ensembling based methods is in discovering a set of microstructural descriptors that cannot be used individually to build a predictive model. The benchmark predictive power (AUC) of the mixed microstructure model is $74.03\%$ compared to around $~52\%$ for a model built on the best individual descriptor. As the materials data science field progresses, we can start utilizing these methods to train models to predict the spatial stress field that evolve due to a complex and collective interaction between crystallographic and geometric parameters.

\section{Conclusions}
\begin{itemize}
    \item Random forest models can predict stress hotspots with 74.03\% AUC in FCC materials under uniaxial tension. The performance of a random forest model trained for all textures in a material is comparable or better than models trained separately for each texture which signals the possibility of common factors causing stress hotspots in a material.
    \item Both texture and geometry derived features contribute to the predictive power of the machine learning model. Grains that become stress hotspots tend to be smaller in size, have high elastic modulus, lower average misorientation and have slip systems more compatible with their neighbors.
    \item The model performance in all the cases can be increased by adding more descriptive features. The geometry based features discussed describe some aspect of the grain and it's nearest neighborhood. Adding long range connectivity based non-crystallographic features might result in an improvement in the model performance.
    \item The feature importances can delineate the microstructural characteristics with the highest impact. Using these insights, machine learning models can be used to design experiments to develop an understanding about the different feature contributions to the target problem.
   \end{itemize}

\subsection{Contributions}
In this article, we have developed a machine learning framework to capture the effects of changing material and texture parameters on stress hotspots. The microstructural features developed in this work can be applied to a range of problems such as prediction of active slip systems for a given texture and loading condition. These methods are also applicable to HEDM obtained datasets.  The feature importance plots are a useful way of determining the most important factors when studying a complex problem with many interacting parameters. As the materials science community moves towards a data driven paradigm, it becomes all the more important to examine these techniques.  



\section{Acknowledgements }
This work was performed at Carnegie Mellon University and has been supported by the United States National Science Foundation award numbers DMR-1307138 and DMR-1507830.

\section{References}

\bibliography{library}

\begin{thebibliography}{53}
\expandafter\ifx\csname natexlab\endcsname\relax\def\natexlab#1{#1}\fi
\providecommand{\url}[1]{\texttt{#1}}
\providecommand{\href}[2]{#2}
\providecommand{\path}[1]{#1}
\providecommand{\DOIprefix}{doi:}
\providecommand{\ArXivprefix}{arXiv:}
\providecommand{\URLprefix}{URL: }
\providecommand{\Pubmedprefix}{pmid:}
\providecommand{\doi}[1]{\href{http://dx.doi.org/#1}{\path{#1}}}
\providecommand{\Pubmed}[1]{\href{pmid:#1}{\path{#1}}}
\providecommand{\bibinfo}[2]{#2}
\ifx\xfnm\relax \def\xfnm[#1]{\unskip,\space#1}\fi
\bibitem[{Anglin et~al.(2014)Anglin, Lebensohn and Rollett}]{Anglin2014}
\bibinfo{author}{Anglin, B.S.}, \bibinfo{author}{Lebensohn, R.A.},
  \bibinfo{author}{Rollett, A.D.}, \bibinfo{year}{2014}.
\newblock \bibinfo{title}{{Validation of a numerical method based on Fast
  Fourier Transforms for heterogeneous thermoelastic materials by comparison
  with analytical solutions}}.
\newblock \bibinfo{journal}{Computational Materials Science}
  \bibinfo{volume}{87}, \bibinfo{pages}{209--217}.
\newblock \DOIprefix\doi{10.1016/j.commatsci.2014.02.027}.
\bibitem[{Aust and Chen(1954)}]{Aust1954}
\bibinfo{author}{Aust, K.T.}, \bibinfo{author}{Chen, N.K.},
  \bibinfo{year}{1954}.
\newblock \bibinfo{title}{{Effect of orientation on the plastic deformation of
  aluminum bicrystals}}.
\newblock \bibinfo{journal}{Acta Metallurgica} \bibinfo{volume}{2},
  \bibinfo{pages}{136--139}.
\bibitem[{Bieler et~al.(2014)Bieler, Eisenlohr, Zhang, Phukan and
  Crimp}]{Bieler2014}
\bibinfo{author}{Bieler, T.R.}, \bibinfo{author}{Eisenlohr, P.},
  \bibinfo{author}{Zhang, C.}, \bibinfo{author}{Phukan, H.J.},
  \bibinfo{author}{Crimp, M.A.}, \bibinfo{year}{2014}.
\newblock \bibinfo{title}{{Grain boundaries and interfaces in slip transfer}}.
\newblock \bibinfo{journal}{Current Opinion in Solid State and Materials
  Science} \bibinfo{volume}{18}, \bibinfo{pages}{212--226}.
\newblock \DOIprefix\doi{10.1016/j.cossms.2014.05.003}.
\bibitem[{Breiman(1996)}]{Breiman1996}
\bibinfo{author}{Breiman, L.}, \bibinfo{year}{1996}.
\newblock \bibinfo{title}{{Out-Of-Bag-Estimation}}.
\newblock \DOIprefix\doi{10.1007/s13398-014-0173-7.2}.
\bibitem[{Breiman(2001)}]{Breiman2001}
\bibinfo{author}{Breiman, L.}, \bibinfo{year}{2001}.
\newblock \bibinfo{title}{{Random Forests}}.
\newblock \bibinfo{journal}{Machine Learning} \bibinfo{volume}{45},
  \bibinfo{pages}{5--32}.
\newblock \DOIprefix\doi{10.1023/A:1010933404324}.
\bibitem[{Bronkhorst(1991)}]{Bronkhorst1991}
\bibinfo{author}{Bronkhorst, C.A.}, \bibinfo{year}{1991}.
\newblock \bibinfo{title}{{Fig. 3.10, Plastic deformation and crystallographic
  texture evolution in face-centered cubic metals}}.
\newblock Ph.D. thesis. Massachusetts Institute of Technology.
\bibitem[{Chen and Guestrin(2016)}]{Chen:2016:XST:2939672.2939785}
\bibinfo{author}{Chen, T.}, \bibinfo{author}{Guestrin, C.},
  \bibinfo{year}{2016}.
\newblock \bibinfo{title}{Xgboost: A scalable tree boosting system}, in:
  \bibinfo{booktitle}{Proceedings of the 22Nd ACM SIGKDD International
  Conference on Knowledge Discovery and Data Mining}, \bibinfo{publisher}{ACM},
  \bibinfo{address}{New York, NY, USA}. pp. \bibinfo{pages}{785--794}.
\newblock \URLprefix \url{http://doi.acm.org/10.1145/2939672.2939785},
  \DOIprefix\doi{10.1145/2939672.2939785}.
\bibitem[{Cohen et~al.(2009)Cohen, Huang, Chen and Benesty}]{Cohen2009}
\bibinfo{author}{Cohen, I.}, \bibinfo{author}{Huang, Y.},
  \bibinfo{author}{Chen, J.}, \bibinfo{author}{Benesty, J.},
  \bibinfo{year}{2009}.
\newblock \bibinfo{title}{Pearson correlation coefficient}, in:
  \bibinfo{booktitle}{Noise Reduction in Speech Processing}.
  \bibinfo{publisher}{Springer}, pp. \bibinfo{pages}{1----4}.
\newblock \DOIprefix\doi{10.1007/978-3-642-00296-0}.
\bibitem[{Counts et~al.(2008)Counts, Braginsky, Battaile and
  Holm}]{counts2008predicting}
\bibinfo{author}{Counts, W.A.}, \bibinfo{author}{Braginsky, M.V.},
  \bibinfo{author}{Battaile, C.C.}, \bibinfo{author}{Holm, E.A.},
  \bibinfo{year}{2008}.
\newblock \bibinfo{title}{Predicting the hall–petch effect in fcc metals
  using non-local crystal plasticity}.
\newblock \bibinfo{journal}{International Journal of Plasticity}
  \bibinfo{volume}{24}, \bibinfo{pages}{1243 -- 1263}.
\newblock \DOIprefix\doi{10.1016/j.ijplas.2007.09.008}.
\bibitem[{Dietterich(2000)}]{dietterich2000ensemble}
\bibinfo{author}{Dietterich, T.G.}, \bibinfo{year}{2000}.
\newblock \bibinfo{title}{Ensemble methods in machine learning}, in:
  \bibinfo{booktitle}{International workshop on multiple classifier systems},
  \bibinfo{organization}{Springer}. pp. \bibinfo{pages}{1--15}.
\bibitem[{Domingos(2012)}]{Domingos2012}
\bibinfo{author}{Domingos, P.}, \bibinfo{year}{2012}.
\newblock \bibinfo{title}{{A few useful things to know about machine
  learning}}.
\newblock \bibinfo{journal}{Communications of the ACM} \bibinfo{volume}{55},
  \bibinfo{pages}{78}.
\newblock \DOIprefix\doi{10.1145/2347736.2347755}.
\bibitem[{Donegan et~al.(2013)Donegan, Tucker, Rollett, Barmak and
  Groeber}]{Donegan2013}
\bibinfo{author}{Donegan, S.P.}, \bibinfo{author}{Tucker, J.C.},
  \bibinfo{author}{Rollett, A.D.}, \bibinfo{author}{Barmak, K.},
  \bibinfo{author}{Groeber, M.}, \bibinfo{year}{2013}.
\newblock \bibinfo{title}{{Extreme value analysis of tail departure from
  log-normality in experimental and simulated grain size distributions}}.
\newblock \bibinfo{journal}{Acta Materialia} \bibinfo{volume}{61},
  \bibinfo{pages}{5595--5604}.
\newblock \DOIprefix\doi{10.1016/j.actamat.2013.06.001}.
\bibitem[{Efron et~al.(2004)Efron, Hastie, Johnstone and
  Tibshirani}]{Efron2004}
\bibinfo{author}{Efron, B.}, \bibinfo{author}{Hastie, T.},
  \bibinfo{author}{Johnstone, I.}, \bibinfo{author}{Tibshirani, R.},
  \bibinfo{year}{2004}.
\newblock \bibinfo{title}{{Least Angle Regression}}.
\newblock \bibinfo{journal}{The Annals of Statistics} \bibinfo{volume}{32},
  \bibinfo{pages}{407--499}.
\newblock \DOIprefix\doi{10.1214/009053604000000067}.
\bibitem[{Fast and Kalidindi(2011)}]{Fast2011}
\bibinfo{author}{Fast, T.}, \bibinfo{author}{Kalidindi, S.R.},
  \bibinfo{year}{2011}.
\newblock \bibinfo{title}{{Formulation and calibration of higher-order elastic
  localization relationships using the MKS approach}}.
\newblock \bibinfo{journal}{Acta Materialia} \bibinfo{volume}{59},
  \bibinfo{pages}{4595--4605}.
\newblock \DOIprefix\doi{10.1016/j.actamat.2011.04.005}.
\bibitem[{Groeber and Jackson(2014)}]{Groeber2014}
\bibinfo{author}{Groeber, M.A.}, \bibinfo{author}{Jackson, M.A.},
  \bibinfo{year}{2014}.
\newblock \bibinfo{title}{{DREAM.3D: A Digital Representation Environment for
  the Analysis of Microstructure in 3D}}.
\newblock \bibinfo{journal}{Integrating Materials and Manufacturing Innovation}
  \bibinfo{volume}{3}, \bibinfo{pages}{5}.
\newblock \DOIprefix\doi{10.1186/2193-9772-3-5}.
\bibitem[{Hamid et~al.(2017)Hamid, Lyu, Schuessler, Wo and
  Zbib}]{Plasticity2017}
\bibinfo{author}{Hamid, M.}, \bibinfo{author}{Lyu, H.},
  \bibinfo{author}{Schuessler, B.J.}, \bibinfo{author}{Wo, P.C.},
  \bibinfo{author}{Zbib, H.M.}, \bibinfo{year}{2017}.
\newblock \bibinfo{title}{{Modeling and Characterization of Grain Boundaries
  and Slip Transmission in Dislocation Density-Based Crystal Plasticity}}.
\newblock \bibinfo{journal}{Crystals} \bibinfo{volume}{7},
  \bibinfo{pages}{152}.
\newblock \DOIprefix\doi{10.3390/cryst7060152}.
\bibitem[{Hanley and McNeil(1982)}]{auc}
\bibinfo{author}{Hanley, J.A.}, \bibinfo{author}{McNeil, B.J.},
  \bibinfo{year}{1982}.
\newblock \bibinfo{title}{{The meaning and use of the area under a receiver
  operating characteristic (ROC) curve.}}
\newblock \bibinfo{journal}{Radiology} \bibinfo{volume}{143},
  \bibinfo{pages}{29--36}.
\newblock \DOIprefix\doi{10.1148/radiology.143.1.7063747}.
\bibitem[{Hull and Rimmer(1959)}]{Rimmer1959}
\bibinfo{author}{Hull, D.}, \bibinfo{author}{Rimmer, D.E.},
  \bibinfo{year}{1959}.
\newblock \bibinfo{title}{{The growth of grain-boundary voids under stress}}.
\newblock \bibinfo{journal}{Philosophical Magazine} \bibinfo{volume}{4},
  \bibinfo{pages}{673--687}.
\newblock \DOIprefix\doi{10.1080/14786435908243264}.
\bibitem[{Johnson and Schuh(2013)}]{JohnsonSchuh2013}
\bibinfo{author}{Johnson, O.K.}, \bibinfo{author}{Schuh, C.A.},
  \bibinfo{year}{2013}.
\newblock \bibinfo{title}{{The uncorrelated triple junction distribution
  function : Towards grain boundary network design}}.
\newblock \bibinfo{journal}{Acta Materialia} \bibinfo{volume}{61},
  \bibinfo{pages}{2863--2873}.
\newblock \DOIprefix\doi{10.1016/j.actamat.2013.01.025}.
\bibitem[{{Kocks, U.F.; Tom{\'{e}}}(1998)}]{KocksTomeWenk}
\bibinfo{author}{{Kocks, U.F.; Tom{\'{e}}}, C.W.H.}, \bibinfo{year}{1998}.
\newblock \bibinfo{title}{{Texture and anisotropy: preferred orientations in
  polycrystals and their effect on materials properties.}}
\newblock \bibinfo{edition}{2} ed., \bibinfo{publisher}{Cambridge university
  press}.
\bibitem[{Kohavi and Others(1995)}]{Kohavi2016}
\bibinfo{author}{Kohavi, R.}, \bibinfo{author}{Others}, \bibinfo{year}{1995}.
\newblock \bibinfo{title}{{A study of cross-validation and bootstrap for
  accuracy estimation and model selection}}, in: \bibinfo{booktitle}{Ijcai},
  \bibinfo{organization}{Montreal, Canada}. pp. \bibinfo{pages}{1137--1145}.
\bibitem[{Lebensohn et~al.(2012)Lebensohn, Kanjarla and
  Eisenlohr}]{Lebensohn2012}
\bibinfo{author}{Lebensohn, R.A.}, \bibinfo{author}{Kanjarla, A.K.},
  \bibinfo{author}{Eisenlohr, P.}, \bibinfo{year}{2012}.
\newblock \bibinfo{title}{{An elasto-viscoplastic formulation based on fast
  Fourier transforms for the prediction of micromechanical fields in
  polycrystalline materials}}.
\newblock \bibinfo{journal}{International Journal of Plasticity}
  \bibinfo{volume}{32-33}, \bibinfo{pages}{59--69}.
\newblock \DOIprefix\doi{10.1016/j.ijplas.2011.12.005}.
\bibitem[{LeCun et~al.(2015)LeCun, Bengio and Hinton}]{LeCun2015}
\bibinfo{author}{LeCun, Y.}, \bibinfo{author}{Bengio, Y.},
  \bibinfo{author}{Hinton, G.}, \bibinfo{year}{2015}.
\newblock \bibinfo{title}{{Deep learning}}.
\newblock \bibinfo{journal}{Nature} \bibinfo{volume}{521},
  \bibinfo{pages}{436--444}.
\newblock \DOIprefix\doi{10.1038/nature14539}.
\bibitem[{Lehockey et~al.(1998)Lehockey, Palumbo and Lin}]{Lehockey19983069}
\bibinfo{author}{Lehockey, E.M.}, \bibinfo{author}{Palumbo, G.},
  \bibinfo{author}{Lin, P.}, \bibinfo{year}{1998}.
\newblock \bibinfo{title}{{Improving the Weldability and Service Performance of
  Nickeland Iron-Based Superalloys by Grain Boundary Engineering}}.
\newblock \bibinfo{journal}{Metallurgical and Materials Transactions A:
  Physical Metallurgy and Materials Science} \bibinfo{volume}{29},
  \bibinfo{pages}{3069--3079}.
\newblock \DOIprefix\doi{10.1007/s11661-998-0214-y}.
\bibitem[{Lienert et~al.(2011)Lienert, Li, Hefferan, Lind, Suter, Bernier,
  Barton, Brandes, Mills, Miller, Jakobsen and Pantleon}]{Lienert2011}
\bibinfo{author}{Lienert, U.}, \bibinfo{author}{Li, S.F.},
  \bibinfo{author}{Hefferan, C.M.}, \bibinfo{author}{Lind, J.},
  \bibinfo{author}{Suter, R.M.}, \bibinfo{author}{Bernier, J.V.},
  \bibinfo{author}{Barton, N.R.}, \bibinfo{author}{Brandes, M.C.},
  \bibinfo{author}{Mills, M.J.}, \bibinfo{author}{Miller, M.P.},
  \bibinfo{author}{Jakobsen, B.}, \bibinfo{author}{Pantleon, W.},
  \bibinfo{year}{2011}.
\newblock \bibinfo{title}{{High-energy diffraction microscopy at the advanced
  photon source}}.
\newblock \bibinfo{journal}{Journal of Materials} \bibinfo{volume}{63},
  \bibinfo{pages}{70--77}.
\newblock \DOIprefix\doi{10.1007/s11837-011-0116-0}.
\bibitem[{Luster and Morris(1995)}]{Luster1995}
\bibinfo{author}{Luster, J.}, \bibinfo{author}{Morris, M.A.},
  \bibinfo{year}{1995}.
\newblock \bibinfo{title}{{Compatibility of deformation in two-phase Ti-Al
  alloys: Dependence on microstructure and orientation relationships}}.
\newblock \bibinfo{journal}{Metallurgical and Materials Transactions A}
  \bibinfo{volume}{26}, \bibinfo{pages}{1745--1756}.
\newblock \DOIprefix\doi{10.1007/BF02670762}.
\bibitem[{Mandal et~al.(2017)Mandal, Gockel, Balachandran, Banerjee and
  Rollett}]{MANDAL201757}
\bibinfo{author}{Mandal, S.}, \bibinfo{author}{Gockel, B.T.},
  \bibinfo{author}{Balachandran, S.}, \bibinfo{author}{Banerjee, D.},
  \bibinfo{author}{Rollett, A.D.}, \bibinfo{year}{2017}.
\newblock \bibinfo{title}{{Simulation of plastic deformation in Ti-5553 alloy
  using a self-consistent viscoplastic model}}.
\newblock \bibinfo{journal}{International Journal of Plasticity}
  \bibinfo{volume}{94}, \bibinfo{pages}{57--73}.
\newblock \DOIprefix\doi{https://doi.org/10.1016/j.ijplas.2017.02.008}.
\bibitem[{Mangal and Holm(2017)}]{Mangal2017c}
\bibinfo{author}{Mangal, A.}, \bibinfo{author}{Holm, E.A.},
  \bibinfo{year}{2017}.
\newblock \bibinfo{title}{{Applied Machine Learning to predict stress hotspots
  II: Hexagonal Close Packed Materials}}.
\bibitem[{Mangal and Holm(2018)}]{mangal2018comparative}
\bibinfo{author}{Mangal, A.}, \bibinfo{author}{Holm, E.A.},
  \bibinfo{year}{2018}.
\newblock \bibinfo{title}{A comparative study of feature selection methods for
  stress hotspot classification in materials}.
\newblock \bibinfo{journal}{arXiv preprint arXiv:1804.09604} .
\bibitem[{Mangal and Kumar(2016)}]{Mangal2017a}
\bibinfo{author}{Mangal, A.}, \bibinfo{author}{Kumar, N.},
  \bibinfo{year}{2016}.
\newblock \bibinfo{title}{{Using big data to enhance the bosch production line
  performance: A Kaggle challenge}}, in: \bibinfo{booktitle}{Proceedings - 2016
  IEEE International Conference on Big Data, Big Data 2016}, pp.
  \bibinfo{pages}{2029--2035}.
\newblock \DOIprefix\doi{10.1109/BigData.2016.7840826},
  \href{http://arxiv.org/abs/1701.00705}{\tt arXiv:1701.00705}.
\bibitem[{Masi et~al.(1980)Masi, Borchi, Gennaro, Borchi, Gennaro, Lakkad,
  Miatt, Parsons, Andrews, Sayers, Kumar, Bai, Dyer, Lord, Gilormini, Yaakob
  and Taib}]{Masi1879}
\bibinfo{author}{Masi, L.}, \bibinfo{author}{Borchi, E.},
  \bibinfo{author}{Gennaro, S.D.}, \bibinfo{author}{Borchi, E.},
  \bibinfo{author}{Gennaro, S.D.}, \bibinfo{author}{Lakkad, S.C.},
  \bibinfo{author}{Miatt, B.B.}, \bibinfo{author}{Parsons, B.},
  \bibinfo{author}{Andrews, K.W.}, \bibinfo{author}{Sayers, C.M.},
  \bibinfo{author}{Kumar, S.S.}, \bibinfo{author}{Bai, V.S.},
  \bibinfo{author}{Dyer, S.R.A.}, \bibinfo{author}{Lord, D.},
  \bibinfo{author}{Gilormini, P.}, \bibinfo{author}{Yaakob, M.K.},
  \bibinfo{author}{Taib, M.F.M.}, \bibinfo{year}{1980}.
\newblock \bibinfo{title}{{Sound velocities and elastic-constant averaging for
  polycrystalline copper}}.
\newblock \bibinfo{journal}{Journal of Physics D: Applied Physics}
  \bibinfo{volume}{13}, \bibinfo{pages}{1879--84}.
\bibitem[{Mercier(2013)}]{mprime}
\bibinfo{author}{Mercier, D.}, \bibinfo{year}{2013}.
\newblock \bibinfo{title}{{Strain Transfer Across Grain Boundaries — Slip
  transfer analysis toolbox 2.0.0 documentation}}.
\newblock \URLprefix \url{http://stabix.readthedocs.io/en/latest/
  slip{\_}transmission.html}.
\bibitem[{Orme et~al.(2016)Orme, Chelladurai, Rampton, Fullwood, Khosravani,
  Miles and Mishra}]{Orme2016}
\bibinfo{author}{Orme, A.D.}, \bibinfo{author}{Chelladurai, I.},
  \bibinfo{author}{Rampton, T.M.}, \bibinfo{author}{Fullwood, D.T.},
  \bibinfo{author}{Khosravani, A.}, \bibinfo{author}{Miles, M.P.},
  \bibinfo{author}{Mishra, R.K.}, \bibinfo{year}{2016}.
\newblock \bibinfo{title}{{Insights into twinning in Mg AZ31 : A combined EBSD
  and machine learning study}}.
\newblock \bibinfo{journal}{Computational Materials Science}
  \bibinfo{volume}{124}, \bibinfo{pages}{353--363}.
\newblock \DOIprefix\doi{10.1016/j.commatsci.2016.08.011}.
\bibitem[{Parker et~al.(2007)Parker, Gunter and Bedo}]{Parker2007}
\bibinfo{author}{Parker, B.J.}, \bibinfo{author}{Gunter, S.},
  \bibinfo{author}{Bedo, J.}, \bibinfo{year}{2007}.
\newblock \bibinfo{title}{{Stratification bias in low signal microarray
  studies}}.
\newblock \bibinfo{journal}{BMC Bioinformatics} \bibinfo{volume}{8},
  \bibinfo{pages}{1--16}.
\newblock \DOIprefix\doi{10.1186/1471-2105-8-326}.
\bibitem[{Pedregosa et~al.(2011)Pedregosa, Varoquaux, Gramfort, Michel,
  Thirion, Grisel, Blondel, Prettenhofer, Weiss, Dubourg and
  Others}]{pedregosa2011scikit}
\bibinfo{author}{Pedregosa, F.}, \bibinfo{author}{Varoquaux, G.},
  \bibinfo{author}{Gramfort, A.}, \bibinfo{author}{Michel, V.},
  \bibinfo{author}{Thirion, B.}, \bibinfo{author}{Grisel, O.},
  \bibinfo{author}{Blondel, M.}, \bibinfo{author}{Prettenhofer, P.},
  \bibinfo{author}{Weiss, R.}, \bibinfo{author}{Dubourg, V.},
  \bibinfo{author}{Others}, \bibinfo{year}{2011}.
\newblock \bibinfo{title}{{Scikit-learn: Machine learning in Python}}.
\newblock \bibinfo{journal}{Journal of Machine Learning Research}
  \bibinfo{volume}{12}, \bibinfo{pages}{2825--2830}.
\bibitem[{Piehler(2009)}]{Piehler2009}
\bibinfo{author}{Piehler, H.R.}, \bibinfo{year}{2009}.
\newblock \bibinfo{title}{{Crystal-Plasticity Fundamentals}}, in:
  \bibinfo{booktitle}{ASM Handbook Volume 22A: Fundamentals of Modeling for
  Metals Processing}. \bibinfo{publisher}{ASM International}, pp.
  \bibinfo{pages}{232--238}.
\bibitem[{Qidwai et~al.(2009)Qidwai, Lewis and Geltmacher}]{Qidwai2009}
\bibinfo{author}{Qidwai, M.A.S.}, \bibinfo{author}{Lewis, A.C.},
  \bibinfo{author}{Geltmacher, A.B.}, \bibinfo{year}{2009}.
\newblock \bibinfo{title}{{Using image-based computational modeling to study
  microstructure – yield correlations in metals}}.
\newblock \bibinfo{journal}{Acta Materialia} \bibinfo{volume}{57},
  \bibinfo{pages}{4233--4247}.
\newblock \DOIprefix\doi{10.1016/j.actamat.2009.05.021}.
\bibitem[{Quinlan(1986)}]{Quinlan1986}
\bibinfo{author}{Quinlan, J.R.}, \bibinfo{year}{1986}.
\newblock \bibinfo{title}{{Induction of Decision Trees}}.
\newblock \bibinfo{journal}{Machine Learning} \bibinfo{volume}{1},
  \bibinfo{pages}{81--106}.
\newblock \DOIprefix\doi{10.1023/A:1022643204877}.
\bibitem[{Quinlan(2014)}]{quinlan2014c4}
\bibinfo{author}{Quinlan, J.R.}, \bibinfo{year}{2014}.
\newblock \bibinfo{title}{{C4. 5: programs for machine learning}}.
\newblock \bibinfo{publisher}{Elsevier}.
\bibitem[{Rajan(2015)}]{Rajan2015}
\bibinfo{author}{Rajan, K.}, \bibinfo{year}{2015}.
\newblock \bibinfo{title}{{Materials Informatics: The Materials " Gene " and
  Big Data}}.
\newblock \bibinfo{journal}{Annu. Rev. Mater. Res} \bibinfo{volume}{45},
  \bibinfo{pages}{153--69}.
\newblock \DOIprefix\doi{10.1146/annurev-matsci-070214-021132}.
\bibitem[{Rathmayr et~al.(2013)Rathmayr, Hohenwarter and Pippan}]{Rathmayr2013}
\bibinfo{author}{Rathmayr, G.B.}, \bibinfo{author}{Hohenwarter, A.},
  \bibinfo{author}{Pippan, R.}, \bibinfo{year}{2013}.
\newblock \bibinfo{title}{{Influence of grain shape and orientation on the
  mechanical properties of high pressure torsion deformed nickel}}.
\newblock \bibinfo{journal}{Materials Science and Engineering A}
  \bibinfo{volume}{560}, \bibinfo{pages}{224--231}.
\newblock \DOIprefix\doi{10.1016/j.msea.2012.09.061}.
\bibitem[{Rhaipu et~al.(2002)Rhaipu, Wise and Bate}]{Rhaipu2002}
\bibinfo{author}{Rhaipu, S.}, \bibinfo{author}{Wise, M.L.H.},
  \bibinfo{author}{Bate, P.S.}, \bibinfo{year}{2002}.
\newblock \bibinfo{title}{{Microstructural Gradients in the Superplastic
  Forming of Ti-6AI-4V}}.
\newblock \bibinfo{journal}{Metallurgical and Materials Transactions,}
  \bibinfo{volume}{33(a)}, \bibinfo{pages}{93--100}.
\newblock \DOIprefix\doi{10.1007/s11661-002-0008-6}.
\bibitem[{Rollett et~al.(2010)Rollett, Lebensohn, Groeber, Choi and
  Li}]{Rollett2010a}
\bibinfo{author}{Rollett, A.D.}, \bibinfo{author}{Lebensohn, R.A.},
  \bibinfo{author}{Groeber, M.}, \bibinfo{author}{Choi, Y.},
  \bibinfo{author}{Li, J.}, \bibinfo{year}{2010}.
\newblock \bibinfo{title}{{Stress hot spots in viscoplastic deformation of
  polycrystals}}.
\newblock \bibinfo{journal}{Modelling and Simulation in Materials Science and
  Engineering} \bibinfo{volume}{18}.
\newblock \DOIprefix\doi{10.1088/0965-0393/18/7/074005}.
\bibitem[{Smith(1948)}]{Smith1948}
\bibinfo{author}{Smith, C.}, \bibinfo{year}{1948}.
\newblock \bibinfo{title}{{Grains, Phases and Interfaces: An interpretation of
  Microstructure}}.
\newblock \bibinfo{journal}{Trans. AIME} \bibinfo{volume}{175},
  \bibinfo{pages}{15--51}.
\bibitem[{Smith(1968)}]{Smith1968}
\bibinfo{author}{Smith, E.}, \bibinfo{year}{1968}.
\newblock \bibinfo{title}{{Cleavage fracture in mild steel}}.
\newblock \bibinfo{journal}{International Journal of Fracture Mechanics}
  \bibinfo{volume}{4}, \bibinfo{pages}{131--145}.
\newblock \DOIprefix\doi{10.1007/BF00188940}.
\bibitem[{Strobl et~al.(2008)Strobl, Boulesteix, Kneib, Augustin and
  Zeileis}]{Strobl2008}
\bibinfo{author}{Strobl, C.}, \bibinfo{author}{Boulesteix, A.L.},
  \bibinfo{author}{Kneib, T.}, \bibinfo{author}{Augustin, T.},
  \bibinfo{author}{Zeileis, A.}, \bibinfo{year}{2008}.
\newblock \bibinfo{title}{{Conditional variable importance for random
  forests.}}
\newblock \bibinfo{journal}{BMC bioinformatics} \bibinfo{volume}{9},
  \bibinfo{pages}{307}.
\newblock \DOIprefix\doi{10.1186/1471-2105-9-307}.
\bibitem[{Taylor(1938)}]{taylor1938plastic}
\bibinfo{author}{Taylor, G.I.}, \bibinfo{year}{1938}.
\newblock \bibinfo{title}{{Plastic strain in metals.}}
\newblock \bibinfo{journal}{J. Inst. Metals} \bibinfo{volume}{62},
  \bibinfo{pages}{307--324}.
\bibitem[{Tibshirani(1996)}]{Tibshirani1996}
\bibinfo{author}{Tibshirani, R.}, \bibinfo{year}{1996}.
\newblock \bibinfo{title}{{Regression Selection and Shrinkage via the Lasso}}.
\newblock \DOIprefix\doi{10.2307/2346178}.
\bibitem[{Varma and Simon(2006)}]{Varma2006}
\bibinfo{author}{Varma, S.}, \bibinfo{author}{Simon, R.}, \bibinfo{year}{2006}.
\newblock \bibinfo{title}{{Bias in error estimation when using cross-validation
  for model selection}}.
\newblock \bibinfo{journal}{BMC Bioinformatics} \bibinfo{volume}{7},
  \bibinfo{pages}{1--8}.
\newblock \DOIprefix\doi{10.1186/1471-2105-7-91}.
\bibitem[{Voce(1955)}]{voce1955practical}
\bibinfo{author}{Voce, E.}, \bibinfo{year}{1955}.
\newblock \bibinfo{title}{{A practical strain-hardening function}}.
\newblock \bibinfo{journal}{Metallurgia} \bibinfo{volume}{51},
  \bibinfo{pages}{219--226}.
\bibitem[{Zare(2015)}]{Zare2015}
\bibinfo{author}{Zare, H.}, \bibinfo{year}{2015}.
\newblock \bibinfo{title}{{FeaLect: Scores Features for Feature Selection}}.
\newblock \URLprefix \url{https://cran.r-project.org/package=FeaLect}.
\bibitem[{Zare et~al.(2013)Zare, Haffari, Gupta and Brinkman}]{Zare2013}
\bibinfo{author}{Zare, H.}, \bibinfo{author}{Haffari, G.},
  \bibinfo{author}{Gupta, A.}, \bibinfo{author}{Brinkman, R.R.},
  \bibinfo{year}{2013}.
\newblock \bibinfo{title}{{Scoring relevancy of features based on combinatorial
  analysis of Lasso with application to lymphoma diagnosis.}}
\newblock \bibinfo{journal}{BMC genomics} \bibinfo{volume}{14},
  \bibinfo{pages}{S14}.
\newblock \DOIprefix\doi{10.1186/1471-2164-14-S1-S14}.
\bibitem[{Zener(1948)}]{zener1948elasticity}
\bibinfo{author}{Zener, C.}, \bibinfo{year}{1948}.
\newblock \bibinfo{title}{Elasticity and anelasticity of metals}.
\newblock \bibinfo{publisher}{University of Chicago press}.

\end{thebibliography}

\setcounter{equation}{0}
\setcounter{figure}{0}
\setcounter{table}{0}
\setcounter{page}{1}
\setcounter{section}{0}
\makeatletter
\renewcommand{\theequation}{S\arabic{equation}}
\renewcommand{\thefigure}{S\arabic{figure}}
\renewcommand{\thetable}{S\arabic{table}}
\renewcommand{\bibnumfmt}[1]{[S#1]}
\renewcommand{\citenumfont}[1]{S#1}

\appendix
\section{Constitutive Parameters: Face Centered Cubic materials} 
\label{AppendixA} 

\begin{figure}[!h]
\centering
\includegraphics[width=0.4\textwidth]{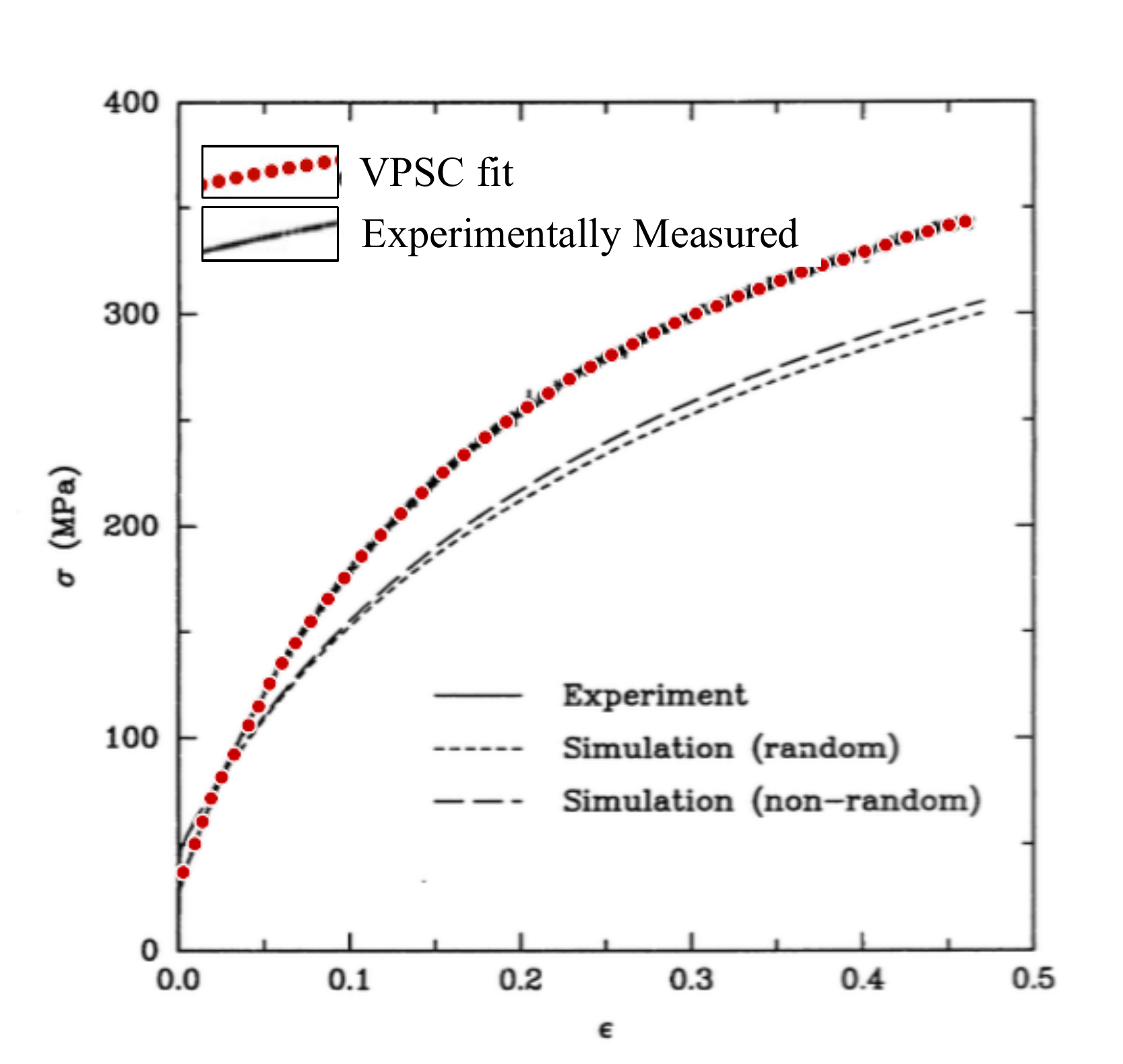}
\caption{VPSC simulation fit to the experimentally observed stress-strain curve for OFHC Copper}
\label{fig:VPSCfit}
\end{figure}

The constitutive model parameters for FCC materials represent oxygen free high thermal conductivity (OFHC) copper from \cite{Masi1879}. The single crystal elastic constants for copper are given in table \ref{HCPelastic1}. FCC materials deform plastically by slip on twelve $ \{111\}<110>$ slip systems. The values of the CRSS and Voce hardening parameters were obtained by fitting the Voce model using the VPSC formulation to an experimentally measured stress-strain curve for uniaxial tension in OFHC copper from \cite{Bronkhorst1991} using the procedure similar to \cite{MANDAL201757}. The results of the fitting are shown in figure \ref{fig:VPSCfit}. The Voce hardening parameters for this hypothetical case are shown in table \ref{Voce}. The boundary conditions correspond to uniaxial tension along Z, with an applied strain rate component along the tensile axis $\dot{\epsilon_{33}} = 1s^{-1}$. The EVPFFT simulation was carried out in 200 steps of 0.01\%, up to a strain of 4\%. 

\begin{figure}[t]
    \centering
    \includegraphics[width=\textwidth]{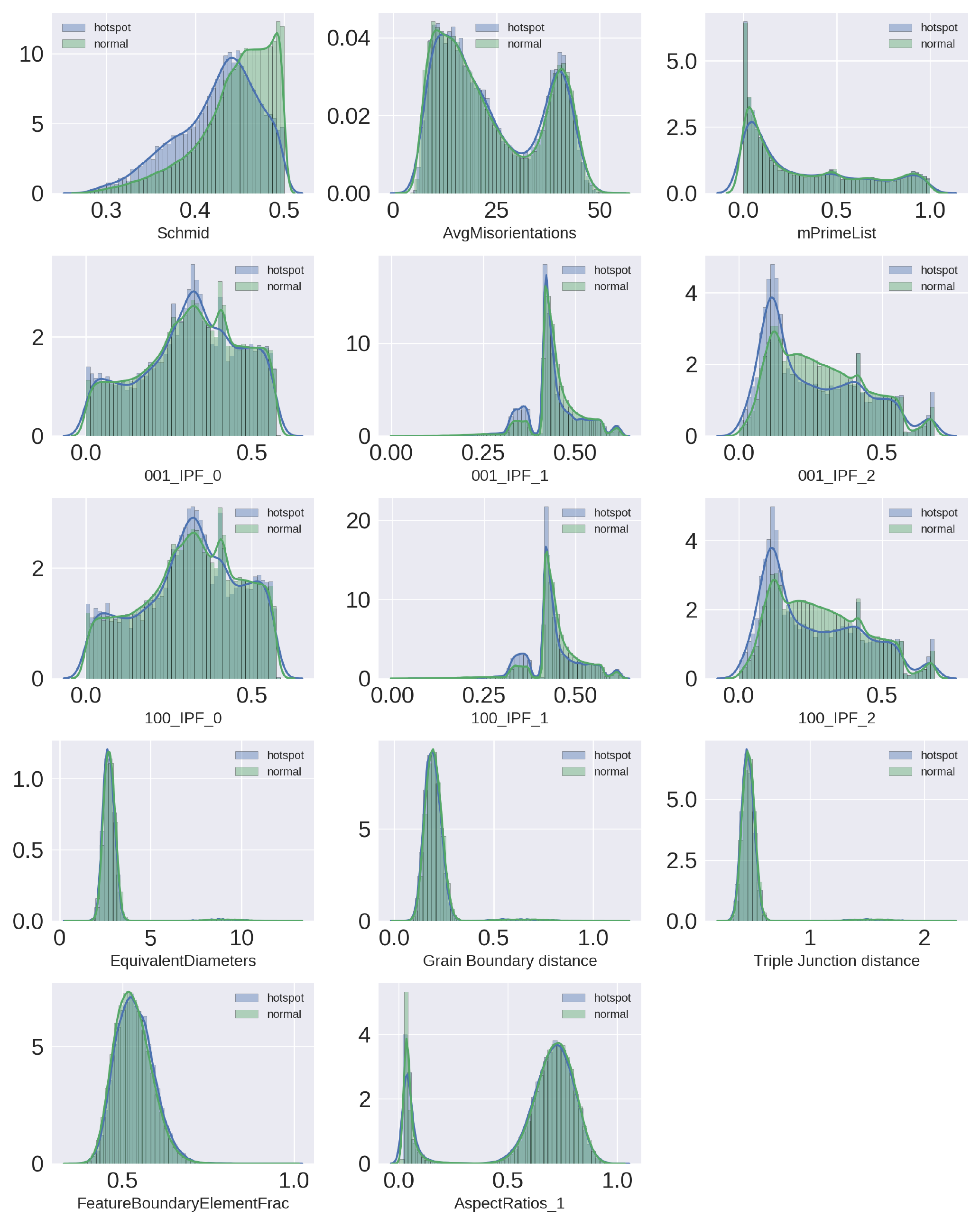}
    \caption{Histogram of the important texture-based features distinguishing hot and normal grains in FCC materials}
    \label{fig:FCC_featurehist}
\end{figure}

To understand how the most predictive features influence hotspot formation, the distribution of these feature values in normal and hot grains are shown in Figure \ref{fig:FCC_featurehist}. The first 3 rows show the texture derived features. The feature distributions for hot and normal grains are different for the all the texture derived features: Schmid factor, the distance of tensile axis (sample Z [001]) from the [111], [110] and [100] crystal directions and the distance of sample X [100] from the [111], [110] and [100] crystal directions. The feature distributions for geometry derived features (bottom 2 rows) are similar, but these features become important in association with texture derived features.\vfill  

\begin{table}[!btp]
\centering
\caption{Single crystal elastic stiffness constants  (in GPa)}
\label{HCPelastic1}
\begin{tabular}{*{4}{c}}
\toprule 
\multicolumn{1}{c}{Material} & \multicolumn{1}{c}{$\mathbf{C_{11}}$} & \multicolumn{1}{c}{$\mathbf{C_{12}}$} & \multicolumn{1}{c}{$\mathbf{C_{44}}$}\\ \midrule
Copper&168.4 &121.4 &75.4\\
\bottomrule
\end{tabular}
\end{table}

\begin{table}[!t]
\centering
\caption{Voce Hardening law parameters imitating Copper}%
\label{Voce}
\begin{tabular}{p{1cm} p{1cm} p{1cm} p{1cm} p{0.8cm}}
\toprule \textbf{CRSS ratio} & $\mathbf{\tau_0^s}$ (MPa) & $\mathbf{\tau_1^s}$ (MPa) & $\mathbf{\theta_0^s}$ & $\mathbf{\theta_1^s}$\\ \midrule
1:1:1 & 7.43 & 102.79 & 356.44 &13.01 \\
\bottomrule
\end{tabular}
\end{table}

\clearpage
\section{Crystallographic and Geometric Descriptors Used for Machine Learning} \label{Table of Acronyms} 

\begin{table}[!h]
\begin{center}
\caption{Feature name descriptions}
\label{Featurenames}
\begin{tabular}{|p{0.18\linewidth}p{0.3\linewidth}|p{0.18\linewidth}p{0.3\linewidth}|}
\toprule 
\textbf{Feature name Abbreviation} & \textbf{Description} & \textbf{Feature name Abbreviation} & \textbf{Description}\\ \midrule
Schmid & FCC Schmid factor & 001\_IPF\_0 & Distance of loading direction from the $<001>$ crystal direction in the  inverse pole figure\\
\midrule
AvgMisorientations & Average misorientation between a grain and its nearest neighbor & 001\_IPF\_1 & Distance of loading direction from the $<110>$ crystal direction in the  inverse pole figure\\
\midrule
Surface Features & 1 if grain touches the periodic boundary else 0 & Surface area volume ratio & Ratio between surface area and volume of a grain\\
\midrule
Omega3s & 3rd invariant of the second-order moment matrix for the grain, without assuming a shape type & Equivalent Diameters & Equivalent spherical diameter of a grain\\
\midrule
mPrimeList & Slip transmission factor for fcc materials & AspectRatios & Ratio of axis lengths (b\/a and c\/a) for best-fit ellipsoid to grain shape\\
\midrule
FeatureVolumes & Volume of grain & QPEuc & Average distance of a grain to quadruple junctions\\
\midrule
TJEuc & Average distance of a grain to triple junctions & NumNeighbors & Number of nearest neighbors of a grain\\
\midrule
GBEuc & Average distance of a grain to grain boundaries & Neighborhoods & Number of grains having their centroid within the 1 multiple of equivalent sphere diameters from each grain\\
\midrule
KernelAvg & Average misorientation within a grain & 001\_IPF\_2 & Distance of loading direction from the $<111>$ crystal direction in the  inverse pole figure \\
\midrule
FeatureBoundary ElementFrac & Fraction of grain touching the periodic boundary & &\\
\bottomrule
\end{tabular}
\end{center}
\end{table}

\clearpage
\section{Cross Validation ROC Curves} \label{CV_ROC} 
\begin{figure}[!h]
\centering
\includegraphics[width=0.8\textwidth]{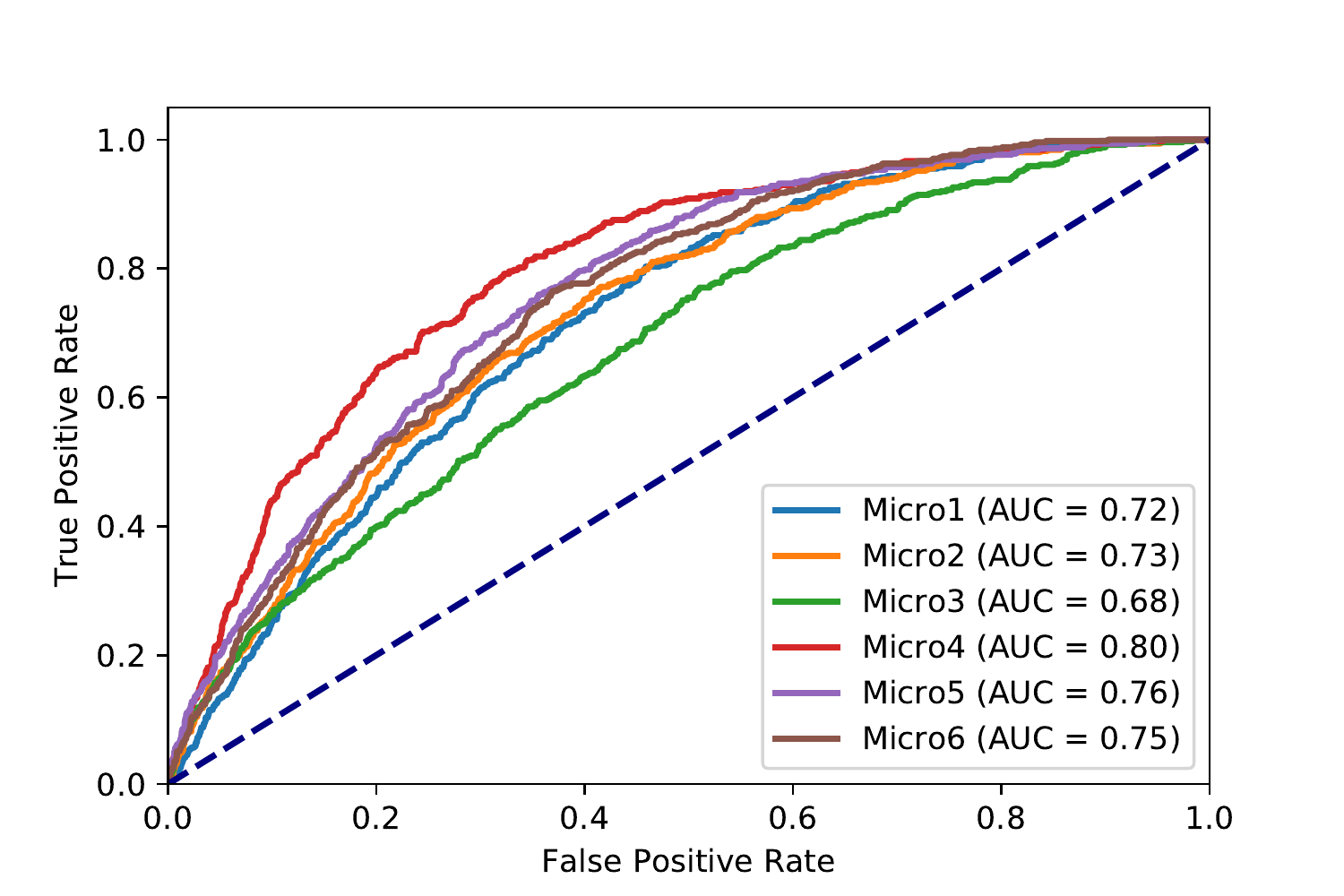}
\caption{ROC curves of the mixed microstructure model for validation microstructures in each representative FCC texture. The AUC is the area under this curve.}
\label{fig:CV_ROC}
\end{figure}

The receiver operating characteristic (ROC) curves for the mixed microstructure random forest model computed using nested cross-validation on representative microstructures are shown in Figure \ref{fig:CV_ROC}. We can see that the mixed microstructure model performs better on some texture classes, but it performs much better than random chance denoted by the dashed blue line.

\end{document}